\documentclass[sigconf]{acmart}

\usepackage{amsmath,amsfonts}

\usepackage{pifont} %use the command \ding

\usepackage{bbding} %对勾叉号
\usepackage{pifont}
\newcommand{\cmark}{\textcolor{green!80!black}{\ding{51}}}
\newcommand{\xmark}{\textcolor{red}{\ding{55}}}

\usepackage{makecell}
\usepackage{colortbl} % table color

\usepackage{lipsum}
\usepackage{tikz}
\usetikzlibrary{arrows.meta}
\usetikzlibrary{automata,positioning}

\usepackage{threeparttable}

\newcommand{\btc}{%
  \leavevmode
  \textcolor{orange}{%
  \rotatebox{345}{% Add rotation here
    \vtop{\offinterlineskip %\bfseries
      \setbox0=\hbox{B}%
      \setbox2=\hbox to\wd0{\hfil\hskip-.03em
      \vrule height .3ex width .15ex\hskip .08em
      \vrule height .3ex width .15ex\hfil}
      \vbox{\copy2\box0}\box2}%
      }
  }%
   \kern-0.5em% Adjust the kerning negative space as needed
}

\newcommand{\brc}{%
  \leavevmode
  \textcolor{magenta}{
    \setbox0=\hbox{\textsf{B}}%
    \dimen0\ht0 \advance\dimen0 0.2ex
    \ooalign{\hfil \box0\hfil\cr
      \hfil\vrule height \dimen0 depth.2ex\hfil\cr}
  }%
  \kern-0.3em% Adjust the kerning negative space as needed
}
\renewcommand*{\arraystretch}{1.5}%
\definecolor{tabred}{RGB}{230,36,0}%
\definecolor{tabgreen}{RGB}{0,116,21}%
\definecolor{taborange}{RGB}{250,124,30}%
\definecolor{tabbrown}{RGB}{171,70,0}%
\definecolor{tabyellow}{RGB}{251,253,169}%
\newcommand*{\vcorr}{%
  \vadjust{\vspace{-\dp\csname @arstrutbox\endcsname}}%
  \global\let\vcorr\relax
}% 

\usepackage{mathtools}

\usepackage{lscape}
\usepackage[figuresright]{rotating}
\usepackage[graphicx]{realboxes}
\usepackage{adjustbox}
\usepackage{caption}

\usepackage{subfigure}

\usepackage{colortbl} 
\usepackage{xfrac}

\usepackage{appendix}
\usepackage{float}

\newcommand{\qw}[1]{\textcolor{blue}{#1}}

\usepackage{fbox} %use box
\usepackage{fancybox}%use other types of boxes
\usepackage{framed}

\usepackage{multirow}

\def\BibTeX{{\rm B\kern-.05em{\sc i\kern-.025em b}\kern-.08em
    T\kern-.1667em\lower.7ex\hbox{E}\kern-.125emX}}
    
\usepackage{CJKutf8}  
\usepackage{pgfplots}
\usepackage{filecontents}

\usepackage{hyperref}

\usepackage{tabularx,makecell, caption}
\usepackage{enumitem} % to control Itemization spacing

\newcolumntype{L}{>{\arraybackslash}X}

\usepackage{multicol}
\usepackage{listings}
\usepackage{blindtext}
\usepackage{etoolbox,xstring,mfirstuc,textcase}
\usepackage{listings}

\usepackage{subfiles}
\usepackage[most]{tcolorbox}

\usepackage{xcolor}
\theoremstyle{plain}

\theoremstyle{definition}

\usepackage{siunitx}
\usepackage{comment}
\usepackage{indentfirst} 
\usepackage{framed} 

\usepackage[font=small,labelfont=bf,tableposition=top]{caption}
\usepackage{booktabs}
\usepackage{dashbox}

\usepackage{listings}
\usepackage{url}

\newenvironment{packeditemize}{
	\begin{list}{$\bullet$}{
			\setlength{\labelwidth}{4pt}
			\setlength{\itemsep}{0pt}
			\setlength{\leftmargin}{\labelwidth}
			\addtolength{\leftmargin}{\labelsep}
			\setlength{\parindent}{0pt}
			\setlength{\listparindent}{\parindent}
			\setlength{\parsep}{0pt}
			\setlength{\topsep}{1pt}}}{\end{list}}

\usepackage{color}
\usepackage{algorithm,algpseudocode} %% select one of them

\renewcommand\footnotetextcopyrightpermission[1]{} % removes footnote with conference information in first column
\begin{document}
\title{BRC20 Snipping Attack}

%=================================================
%author
%=================================================

%\begin{comment} 

\author{Minfeng Qi$^{1,\textcolor{green}{*}}$, Qin Wang$^{2,\textcolor{green}{*}}$, Ningran Li$^{3}$, Shiping Chen$^{2}$, Tianqing Zhu$^{1}$}
\thanks{$^{\textcolor{green}{*}}$ Equal contribution}%Equal contribution} effective when adding additional authors
\affiliation{
\textit{$^1$City University of Macau}, China  \\
\textit{$^2$CSIRO Data61 $|$ $^3$University of Adeleide}, Australia  
}

%\end{comment}

%=================================================
%abstract
%=================================================

\begin{abstract}

In this paper, we introduce and implement \textit{BRC20 sniping attack}. Our attack manipulates the BRC20 token transfers in open markets and disrupts the fairness among bidding participants. The long-standing principle of "highest bidder wins" is rendered ineffective.

Typically, open BRC20 token markets rely on Partially Signed Bitcoin Transactions (PSBT) to broadcast selling intents and wait for buying auctions. Our attack targets the BRC20 buying process (i.e., transfer) by injecting a front-running transaction to complete the full signature of the PSBT. At its core, the attack exploits the mempool’s fee-based transaction selection mechanism to snipe the victim transaction, replicate metadata, and front-run the legesmate transction. This attack applies to platforms using PSBT for BRC20 token transfers, including popular Bitcoin exchanges and marketplaces (e.g., Magic Eden, Unisat, Gate.io, OKX).

We implemented and tested the attack on a Bitcoin testnet (regtest), validating its effectiveness through multiple experimental rounds. Results show that the attacker consistently replaces legitimate transactions by submitting higher-fee PSBTs. We have also made responsible disclosures to the mentioned exchanges.

%In this paper, we propose and implement the \textit{BRC20 Sniping Attack}, a new attack targeting the BRC20 token/inscription transfer process in the Bitcoin network. The attack exploits the mempool’s fee-based transaction selection mechanism, where attackers can front-run legitimate transactions by outbidding the transaction fee. This leads to a situation where attackers can hijack BRC20 token transfers, disrupting the fairness of token markets and discouraging legitimate participants. 

%Our attack can be applied to platforms that use PSBT (Partially Signed Bitcoin Transactions) for BRC20 token transfers, including popular Bitcoin exchanges and marketplaces (i.e., Magic Eden). We implement and evaluate the attack locally on a Bitcoin testnet (regtest), confirming its effectiveness across multiple experimental rounds. Experimental results demonstrate that the attack consistently allows the attacker to replace the legitimate transaction by submitting a higher-fee PSBT. To our knowledge, this is the first time a \textit{BRC20 Sniping Attack} has been formally proposed and tested, shedding light on the vulnerabilities of PSBT-based BRC20 transfers in the Bitcoin ecosystem.

\end{abstract}

%\keywords{BRC20, PSBT, Transaction fee, Attack, Mempool, Bitcoin}

%=================================================
\maketitle
%=================================================   

%=================================================   
\section{Introduction}
%=================================================   

Since mid-2024, the Bitcoin network has witnessed~\cite{chain2024report}
a sharp rise in so-called \textit{mempool snipping}, a fee-based form of transaction front-running that threatens to disrupt ordinary users’ ability to participate in Ordinals~\cite{ordinaltheory2024} and BRC20 trading~\cite{binance1,binance2,binance3,binance4,binance5,binance6}. At its core, mempool snipping relies on the attacker’s capacity to monitor unconfirmed transactions and outbid legitimate buyers by submitting higher-fee alternatives. No direct loss of funds is reported for users who experience this phenomenon, while the damage manifests in a degraded marketplace. Participants may be repeatedly outbid and ultimately deterred from engaging in token purchases.

According to data shared by community members, this practice is no longer a rare occurrence. The Wizards of Ord “Snipes” Discord channel~\cite{snipe2024discord} recorded 220 daily snipes in one instance, amounting to 3.9 BTC (approximately \$247,000), and similar activity levels have been reported consistently over the course of several weeks. High-profile figures have encountered similar setbacks. For instance, on July 10, an individual affiliated with the Bored Ape Yacht Club engaged in what community members termed a “sniping battle,” ultimately forfeiting over \$5{,}800 in fees without succeeding in the attempted purchase of an Ape Hoodie Ordinal~\cite{snip2024battle}. Another indicative incident occurred during the highly anticipated OrdiBots mint~\cite{phoenix2024ordibots}, which had garnered substantial attention among Ordinals enthusiasts. The enthusiasm quickly soured when numerous participants experienced mempool sniping throughout the launch.

% Faced with users losing out on the tokens they intended to purchase, Magic Eden~\cite{magic2024}, a prominent marketplace, publicly acknowledged the sniping issues while the OrdiBots team vowed to compensate affected users via asset airdrops. 

Despite widespread awareness of mempool sniping, the technical underpinnings of Ordinals and BRC20 transfers have yet to be fully examined in the context of fee-driven front-running. At the heart of these trading workflows lies the partially signed Bitcoin transaction (PSBT) mechanism, which enables sellers to broadcast sale data and buyers to finalize the purchase. This convenience, however, introduces an opening for adversaries to intercept, modify, or outbid legitimate transactions before they reach confirmation, a phenomenon we term the \textit{BRC20 snipping attack}. 

\vspace{-0.1em}
\begin{center}
\fbox{%
    \begin{minipage}{0.85\linewidth}
        \textbf{Evidence:}  A poll~\cite{poll2024snip} conducted by Magisat’s founder on June 12 revealed that, out of 229 respondents, \textbf{73\%} confirmed having been sniped at least once while attempting to secure Ordinal assets.
    \end{minipage}
} 
\end{center}
\vspace{-0.1em}

In this paper, we present a deep dive into \textit{how BRC20 token transfers via PSBT are susceptible to sniping.}

\ding{172} \textbf{We conduct the first formal treatment of the PSBT-based BRC20 transfer process, which is the key process for building open marketplaces within the Bitcoin ecosystem.}

BRC20 tokens, unlike Ethereum-based tokens~\cite{erc20,erc721,wang2021non}, lack native smart contract support. Instead, BRC20s rely on Bitcoin's UTXO (Unspent Transaction Output) model. This necessitates a structured transfer mechanism to handle token movements while preserving the system integrity. The PSBT standard (BIP-174~\cite{bip174}) provides a means to separate transaction creation and signing across multiple parties, making it a cornerstone for implementing BRC20 token marketplaces. However, this introduces vulnerabilities as incomplete transactions are visible and manipulable before their finalization.

Our analysis formalizes the BRC20 transfer workflow through PSBT by decomposing its phases: seller initialization, buyer signing, and network broadcast. We introduce a mathematical model capturing the dependencies between transaction inputs, outputs, and the metadata inscribed in PSBTs. By detailing how token transfer inscriptions (e.g., JSON-based metadata) are integrated into the Bitcoin blockchain, we identify the system's reliance on mempool transparency and fee-driven prioritization.

\ding{173} \textbf{We propose a new attack we term \textit{BRC20 snipping attact}, where partial/incompleted PSBTs can be replicated with a higher fee, displacing legitimate transactions within the mempool from malicious buyers.}

Our sniping attack exploits the transparent nature of the mempool and the predictable workflow of PSBT-based transactions. In our attack, the adversary monitors the mempool for unconfirmed BRC20 transactions, identifies partially signed PSBTs containing token transfer instructions, and crafts a competing transaction. By replicating the legitimate transaction's core details and outbidding it with a higher fee, the attacker ensures their transaction is prioritized for block inclusion, effectively invalidating the original.

Our attack not only disrupts the fairness of BRC20 marketplaces but also undermines user confidence in their reliability. More importantly, the damage can amplify significantly as our attack is widely applicable (details in Table~\ref{tab:brc-tokens}).

It is important to note that we do \textbf{NOT} claim novelty in the concept of mempool sniping, as extensively discussed in prior works. However, our study is the first to investigate the sniping attack specifically within the context of BRC20 token transfers and the PSBT protocol. The complexities introduced by these additional protocols pose greater analytical challenges compared to simpler mempool injection or traditional front-running attacks. Our attack, despite such complexity, demonstrates a fully operational implementation. For better understanding, we also summarize relevant (mempool) attack vectors in Table~\ref{tab:attack_comparison}.

\ding{174} \textbf{We implement and validata our attack in a controlled environment. We conduct a detailed local experiment that simulates real-world conditions of network congestion, fee competition, and opportunistic adversaries.}

We developed a controlled environment using Bitcoin's regtest mode. We emulate real-world scenarios, including mempool congestion, varying transaction fees, and adversarial behaviors. We created a local marketplace for BRC20 tokens and implemented the attack by crafting sniping transactions with varying fee levels.

Our experiments show that attackers with minimal resources can consistently outbid legitimate buyers, achieving block inclusion for their malicious transactions. We measured the success rate of the attack under different network conditions and fee strategies, confirming its reliability. Additionally, we tracked the impact on token balances, demonstrating how the attacker successfully redirected tokens to their control while rendering the legitimate transaction invalid. These findings validate the attack's practicality.

\ding{175} \textbf{We present our suggestion for mitigation}.

We propose three mitigation strategies: (i) adding mempool protection, (ii) enabling dynamic fee escalation, and (iii) implementing a fee-locking mechanism. For instance, the advanced fee-locking mechanism embeds a maximum fee commitment within the PSBT, preventing attackers from arbitrarily escalating fees. This mechanism leverages cryptographic commitments to ensure that only transactions adhering to the agreed-upon fee structure are accepted by the network. We also explore the feasibility of integrating dynamic fee adjustments, where users can pre-authorize higher fees in case their initial transaction is outbid.

\smallskip
In short, our contributions are:

\begin{packeditemize}
   \item formalisation for the PSBT-based BRC20 transfer process (\S\ref{sec-formal});
    \item the new BRC20 sniping attack (\S\ref{sec-attack}, \S\ref{sec-applicable});
    \item implementation and validation of the attack (\S\ref{sec-imple});
    \item mitigation strategies (\S\ref{sec-mitigation}).

\end{packeditemize}

\begin{table*}[!ht]
\centering
\caption{\textcolor{teal}{Mempool}-Related Attacks / Exploits}
\label{tab:attack_comparison}
\vspace{-0.2em}
\renewcommand{\arraystretch}{1.1}
\resizebox{\textwidth}{!}{%
\begin{tabular}{c|cc|cc|c}

\midrule

\multicolumn{1}{c}{\textbf{Aspect}} & 
\multicolumn{1}{c}{\textbf{Target}}  & 
\multicolumn{1}{c}{\textbf{Objective}} &
\multicolumn{1}{c}{\textbf{Key Exploit}} & 
\multicolumn{1}{c}{\textbf{Mechanism}}  & 
\multicolumn{1}{c}{\textbf{Platform}}   
\\ 

\midrule

Frontrunning~\cite{daian2020flash,zhou2021high,zhang2022frontrunning,wang2022exploring}  & Pending transactions  & Displace transaction & Fee bidding & Fee escalation or reduction (backrunning)  & EVM-compatible    \\ 

MEV arbitrage~\cite{eskandari2020sok,heimbach2022sok,yang2024sok} & Transaction orderings  & Extract unfair value & Miner prioritization & Order manipulation/censorship  &  EVM-compatible \\ 

Sniping bots~\cite{cernera2023token,cernera2023ready,cernera2024warfare}  & Token/NFT auctions  & Capture mints  & Gas auctions & Automated gas bidding in mints    & EVM-compatible  \\  

Imitation attack~\cite{qin2023blockchain} & Pending tx/contracts   & Extract contract value  & Contract cloning & Copy and frontrun (see above) contracts  & EVM-compatible \\ 

\cmidrule{1-2}

RBF exploits~\cite{li2023transaction,qi2024brc20,txpinHTLC} & Low-fee transactions    & Accelerate confirmation  & Fee prioritization 
 & Replace transactions & Bitcoin (UTXO) \\ 

Tx malleability~\cite{decker2014bitcoin,andrychowicz2015malleability} & Transaction IDs & Invalidate/hijack Tx & Signature alteration & Alter tx IDs before confirmation & Bitcoin (UTXO) \\

Pinning attacks~\cite{qi2024brc20,bicoinop24,txpinHTLC}  & Inscriptions/BRC20s & Drain out liquidity  & Locking UTXOs & Fee escalation + full withdraw order  & Bitcoin (UTXO) \\  

\cmidrule{1-2}

Timejacking attack~\cite{zhang2023time}  & Node timestamps & Disrupt transaction order & False timestamp & Alter node time to influence validation & All blockchains \\ 

Double-spend~\cite{negy2020selfish,feng2019selfish,eyal2018majority} & UTXOs & Spend coins twice & Conflicting Tx/branch & Broadcast multiple tx with same inputs & All blockchains \\ 

DDoS~\cite{wu2020survive,saad2019shocking,saad2019mempool,vasek2014empirical} & Mempool capacity & Block valid tx propagation & Spam tx flooding & Overload mempool with low-priority tx & All blockchains \\

\midrule

\textbf{Ours}  & \textbf{PSBT/Inscription/BRC20} & \textbf{Hijack token transfers}   & 
\textbf{Metadata in PSBT} & \textbf{Fee escalation + PSBT metadata replication}      & \textbf{Bitcoin (UTXO)   } \\ 

\bottomrule
\end{tabular}%
}
\end{table*}

\begin{center}
%\fbox{%
   \colorbox{teal!10}{
    \begin{minipage}{0.85\linewidth}
        \textbf{Responsible disclosure.} We have notified our attack findings to four leading BRC20 token platforms and marketplaces, i.e., \textit{Magic Eden}, \textit{Unisat}, \textit{Gate.io}, \textit{OKX},  which are potentially affected by this vulnerability. 
        \end{minipage}

} 
%}
\end{center}

%================================================= 
\section{Technical Warmups}
%================================================= 

\smallskip
\noindent\textbf{Bitcoin's UTXO.} Bitcoin's UTXO (Unspent Transaction Output)~\cite{nakamoto2008bitcoin,garay2024bitcoin} is a fundamental concept in its transaction model, representing outputs of transactions that have not yet been spent. Each UTXO is tied to a Bitcoin address and controlled by a private key, allowing the holder to spend it. UTXOs are indivisible, meaning they must be fully consumed in a transaction, with any leftover value returned as a new UTXO to the sender. The Bitcoin network maintains a global UTXO set, which tracks all unspent outputs and forms the basis for validating transactions. A transaction spends UTXOs as inputs and creates new UTXOs as outputs, ensuring a clear lineage for every Bitcoin. We present a formalized version in $\S\ref{sec-formal}$.

The UTXO model enhances security, efficiency, and transparency. UTXOs prevent double-spending since each can only be spent once and ensure traceability through their transactional history. Unlike the account model used in blockchains like Ethereum~\cite{wood2014ethereum,wang2023account}, UTXOs eliminate the need to manage balances directly, instead relying on the creation and consumption of outputs. 

\begin{figure}[!h]
    \centering
    \includegraphics[width=0.9\linewidth]{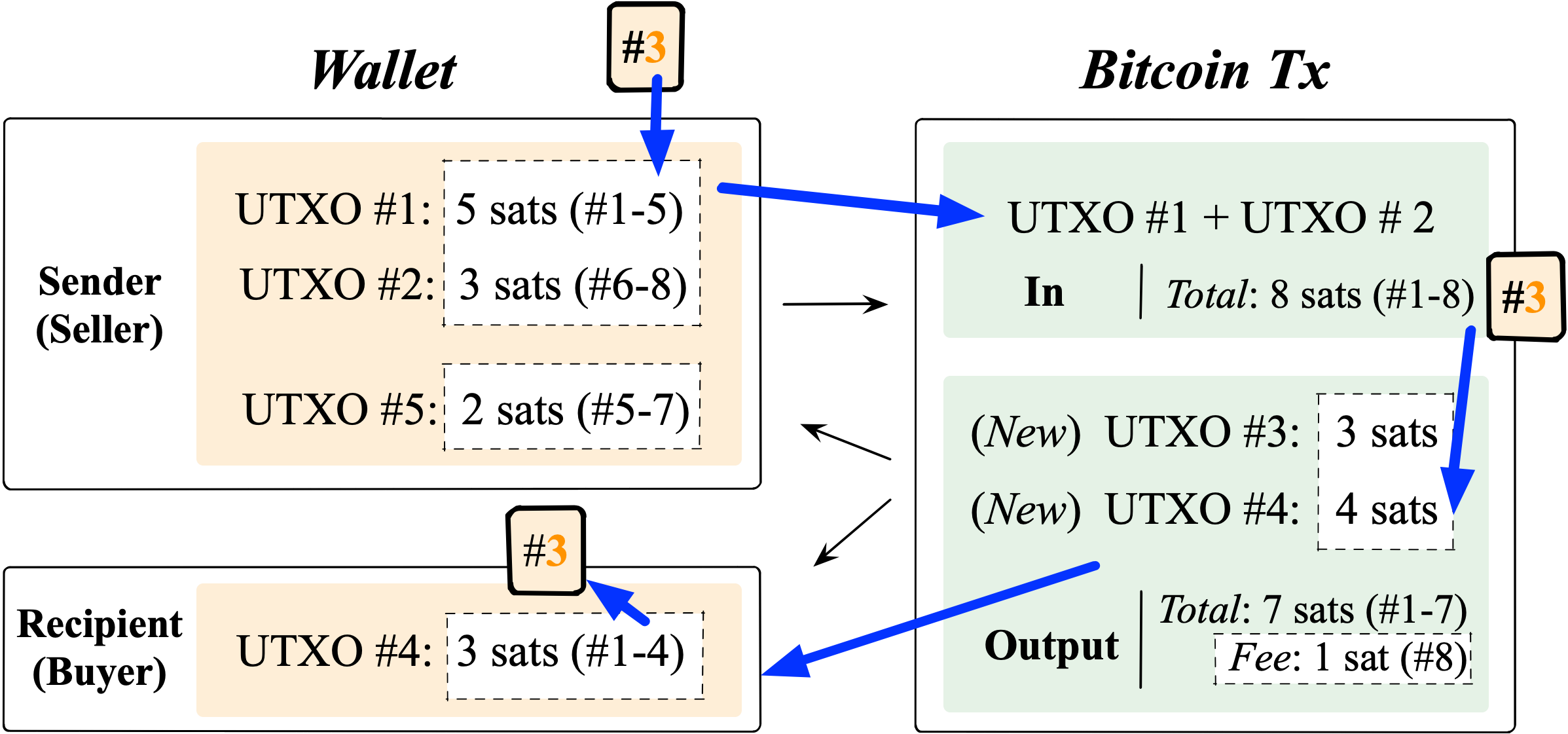}
    \caption{Transferring BRC20/Inscriptions via the UTXO model}
    \label{fig:transfer}
   \vspace{-0.15in}
\end{figure}

\smallskip
\noindent\textbf{Bitcoin's inscription.} 
Creating an inscription transaction involves two distinct phases~\cite{wang2023understanding,qi2024brc20}:

\begin{packeditemize}
    \item \textbf{Commit transaction.} This phase begins with the creation of a taproot output~\cite{bip341,bip342} that cryptographically commits to a script referencing the inscription data. Importantly, this commitment is designed to reference the data without directly revealing its content, ensuring confidentiality until the appropriate stage. The commit transaction is then broadcast to the Bitcoin network, where it gets included in a block on the blockchain. This process securely anchors the commitment on-chain, forming the foundation for the subsequent phase.

    \item \textbf{Reveal transaction.} In this phase, the output from the commit transaction is spent in a new transaction known as the reveal transaction. This transaction includes the actual inscription data embedded within its script, fully disclosing the committed content. Once the reveal transaction is confirmed on the blockchain, the inscription data becomes permanently stored on-chain, ensuring public accessibility, verifiability, and immutability. This phase completes the inscription process, allowing the data to be associated with the Bitcoin network permanently.
\end{packeditemize}

\smallskip
\noindent\textbf{BRC20.} 
The BRC20 protocol~\cite{binance2,binance3,wang2023understanding} builds upon inscription technology to embed personalized content into each unique satoshi, the smallest unit of Bitcoin. The inscribed data is assigned a P2TR (Pay-to-Taproot) script type and stored within the SegWit witness section~\cite{bip141} of a transaction, ensuring compatibility with the Bitcoin network. Inscriptions can take diverse formats, including JSON-based text, images, audio, and even highly compressed video files. For text-based BRC20 tokens, such as ORDI \cite{ordicoin}, the on-chain inscription includes key details: the protocol name ($\mathsf{brc20}$), operation type (e.g., $\mathsf{transfer}$), token name (e.g., $\mathsf{ordi}$), total token supply ($\mathsf{max}$), maximum tokens minted per round ($\mathsf{lim}$), and the quantity to be minted or transferred ($\mathsf{amt}$). This structured approach ensures accurate and permanent recording of token metadata on-chain.

The protocol leverages Ordinal Theory \cite{ordinaltheory2024}, which functions as a numbering system to index every indivisible satoshi, enabling them to be uniquely tracked and transferred as individual entities. This indexing mechanism ensures that each satoshi carrying an inscription, such as BRC20 metadata, becomes non-fungible, effectively differentiating it from others. This contrasts with Ethereum, which inherently supports smart contracts for token creation and management, making the BRC20 protocol a distinct and innovative approach to tokenization within the Bitcoin ecosystem.

\smallskip
\noindent\textbf{Transferring BRC20.} BRC20 tokens lack built-in smart contracts and instead operate using Bitcoin's UTXO model and transaction inscriptions. Inscriptions only store plaintext messages about BRC-20 actions, not executable logic. Real-time balance updates depend on off-chain indexers for data retrieval. We explain more below.

\begin{packeditemize}
    
\item\textbf{On-chain storage.} 
BRC-20 tokens are fully recorded on-chain, with their movements tied to Bitcoin’s native transactions. The \texttt{Transfer} operation corresponds to the sending and receiving of satoshis on the layer-one Bitcoin network. However, the Bitcoin network only logs transaction data, not the real-time status of BRC-20 tokens or balances. Unlike account-based systems, advanced features like balance tracking are offloaded to external layers, keeping the on-chain ledger lightweight and efficient.

\item\textbf{Off-chain retrievals.} 
Off-chain indexers~\cite{wang2023understanding, wang2024bridging} handle real-time data for BRC-20 tokens. These services track token minting, determine the cessation points, and map token trades to specific wallets. They only observe on-chain transactions without modifying them, ensuring data integrity. If an indexer fails, token data remains secure, as it can be reconstructed by reprocessing blockchain transactions. Many wallets, like Ordinals Wallet~\cite{ordinalwallet}, UniSat~\cite{unisat}, and Binance’s hot wallet~\cite{binanceweb3wallet}, integrate indexer functionalities seamlessly for users.

\end{packeditemize}

%================================================= 
\section{Formalising BRC20 Transfer on PSBT}
\label{sec-formal}
%================================================= 

\subsection{Basics}

\noindent\textbf{Transaction model.}
In the Bitcoin network, transactions are the fundamental units\footnote{We denote Bitcoin unit as \btc~\,(i.e., BTC) and BRC20 token unit as \brc~.} that facilitate the transfer of value, operating under the UTXO model. A Bitcoin transaction \( \mathcal{T} \) can be formally defined as a tuple \( \mathcal{T} = (\text{txid}, \text{In}, \text{Out}, \text{Witness}) \), where:
\begin{packeditemize}
    \item \( \textbf{txid} = \mathcal{H}(\text{serialize}(\mathcal{T})) \) is a unique identifier for transactions, derived from the double SHA-256 hash function \( \mathcal{H} \) applied to the serialized transaction data.
    \item \( \mathbf{In} = \{ I_1, I_2, \ldots, I_n \} \) is the set of inputs. Each input \( I_i \) is a tuple:
    \(
    I_i = (\textbf{txid}^{(i)}, \textbf{vout}^{(i)}, \text{scriptSig}^{(i)}, \text{sequence}^{(i)})
    \)
    where \( \textbf{txid}^{(i)} \) and \( \textbf{vout}^{(i)} \) reference the originating transaction and the specific output index being consumed, respectively. The \( \text{scriptSig}^{(i)} \) provides the necessary cryptographic proof to authorize the spending of the referenced UTXO. \(\text{sequence}^{(i)} \) enables features such as \textit{Replace-By-Fee} (RBF, to be explained soon) and facilitating relative lock-time constraints.
    
    \item \(\mathbf{Out}=\{O_1, O_2, \ldots, O_m\}\) is the set of outputs. Each output \( O_j \) is defined as:
    \(
    O_j = (\text{amount}_j, \text{scriptPubKey}_j)
    \)
    where \( \text{amount}_j \) specifies the value being transferred, and \( \text{scriptPubKey}_j \) contains the locking script that sets the conditions under which the output can be spent in future transactions. There are several commonly used output types, including \textit{P2PKH (Pay-to-PubKey-Hash)}, \textit{P2SH (Pay-to-Script-Hash)}, \textit{P2WPKH (Pay-to-Witness-PubKey-Hash)}, \textit{P2WSH (Pay-to-Witness-Script-Hash)}, and \textit{P2TR (Pay-to-Taproot)}.
    
    \item \( \mathbf{Witness} \) is the data structure introduced by Segregated Witness (SegWit) to hold the unlocking data for witness-based outputs like \textit{P2WPKH}, \textit{P2WSH}, and also \textit{P2TR}. Conceptually, the witness serves as the new location for the data that would otherwise appear in the \( \text{scriptSig} \) of legacy transactions. Unlike \( \text{scriptSig} \), the witness section does not employ Bitcoin script opcodes directly; rather, it is composed solely of data pushes. Each data push corresponds to a piece of unlocking data (e.g., signatures, public keys, or entire witness scripts) required to satisfy the output’s conditions.

\end{packeditemize}

\smallskip
\noindent\textbf{UTXO.}
The UTXO set \( \mathcal{U} \) is a finite collection of all unspent outputs \( ( \textbf{txid}, \textbf{vout}, \text{amount}, \text{scriptPubKey} ) \). The state transition equation below illustrates how \( \mathcal{U} \) evolves with each transaction \( \mathcal{T} \). 

\begin{align*}
\mathcal{U}' &= \mathcal{U} \setminus \left\{ \left( \textbf{txid}^{(i)}, \textbf{vout}^{(i)} \right) \mid I_i \in \mathbf{In} \right\} \\
&\quad \cup \left\{ \left( \textbf{txid}, j, \text{amount}_j, \text{scriptPubKey}_j \right) \mid O_j \in \mathbf{Out} \right\}.
\end{align*}

When a transaction $\mathcal{T}$ is processed, it consumes specific UTXOs referenced by its inputs  $\mathbf{In}$. These consumed UTXOs are removed from the current UTXO set $\mathcal{U}$. Simultaneously, the transaction generates new UTXOs through its outputs $\mathbf{Out}$. These new UTXOs are added to form the updated UTXO set $\mathcal{U'}$.

% \qw{another good reference Sec2 at \url{https://eprint.iacr.org/2020/476.pdf}} \mf{referred} 

%--------------------------------
\subsection{Transaction Replacement}

\noindent\textbf{Transaction Fee.}
Transaction fees serve as an incentive for miners to include transactions in the next block and act as a mechanism to prevent network abuse by adding a small cost to each transaction. The transaction fee (\(f\)) is determined by the difference between the input and output values, which can be calculated as:
\begin{equation}
\label{tx-fee}
   f = \sum_{i=1}^{n} \text{amount}(I_i) - \sum_{j=1}^{m} \text{amount}(O_j). 
\end{equation}

Miners collect this difference as compensation for processing the transaction and securing the network. The fee is typically measured in satoshis per byte (sat/byte), and the exact fee needed for a transaction depends on network congestion and competition for block space at the time. Fig.\ref{fig:tx-fee} depicts the average Bitcoin transaction fee in USD (left y-axis) and the number of Bitcoin transactions per day (right y-axis) over time, spanning from January 2022 to October 2024. The transaction fee remained relatively low and stable, fluctuating around \$20 until early 2023. After this point, there was a noticeable increase in volatility, with transaction fees spiking multiple times, reaching above \$120 at several points in 2024. The creation of the BRC20 standard and the Bitcoin halving event appear to be associated with a significant spike in both the average transaction fee and the number of transactions.

% In Bitcoin Core, the minimum relay transaction fee is set by the $\mathsf{minrelaytxfee}$ parameter, with a default of $0.00001$ BTC per kilobyte~\cite{}. Transactions below this threshold are considered free, but are only forwarded if space is available in the mempool, otherwise, they are discarded.

\smallskip
\noindent\textbf{Transaction replacement.} The transaction replacement mechanism allows users to replace an unconfirmed transaction (\( i.e., \mathcal{T} \)) with a new one (\(i.e., \mathcal{T}' \)) that offers a higher fee. It mandates that the input sets remain identical, 
\(
\mathbf{In}' = \mathbf{In},
\)
ensuring that both transactions consume the exact same UTXOs from the current UTXO set \( \mathcal{U} \). However, the outputs \( \mathbf{Out}' \) in \( \mathcal{T}' \) are required to differ from those in \( \mathcal{T} \). Specifically, \( \mathcal{T}' \) typically lowers the output amounts to assign a higher transaction fee, which can be mathematically expressed as:
\[
\sum_{j=1}^{m'} \text{amount}(O_j') < \sum_{j=1}^{m} \text{amount}(O_j),
\]
where \( m \) and \( m' \) denote the number of outputs in \( \mathcal{T} \) and \( \mathcal{T}' \), respectively. The increased fee \( f' = \sum_{i=1}^{n} \text{amount}(I_i) - \sum_{j=1}^{m'} \text{amount}(O_j') \) in \( \mathcal{T}' \) serves as an incentive for miners to prioritize the replacement transaction over the original one.

% Another critical aspect of RBF is the manipulation of the sequence numbers \( \textit{sequence}^{(i)} \) in the inputs. For a transaction to signal its eligibility for replacement, the sequence numbers must be set below the maximum value, typically:
% \(
% \textit{sequence}^{(i)} < \textit{0xFFFFFFFF},
% \)
% for all \( I_i \in \mathbf{In} \). This modification indicates that the transaction is replaceable, allowing it \( \mathcal{T}' \) to supersede \( \mathcal{T} \) in the mempool. 

% \qw{extract link from this example \url{https://x.com/0xBongo/article/1807650260736745616/media/1807607519554207744}}
\begin{center}
\fbox{%
    \begin{minipage}{0.85\linewidth}
        \textbf{Example}: The Bitcoin history  (cf. \href{https://x.com/0xBongo/article/1807650260736745616/media/1807607519554207744}{link}) shows that a transaction (i.e., \textcolor{teal}{txid.fc7c}) was replaced with a new transaction (i.e., \textcolor{red}{txid.e744}) with a fee rate of 12.8 sat/vB, higher than the initial 9.04 sat/vB.
    \end{minipage}
}
\end{center}

% \smallskip
% \noindent\textbf{\qw{PSBT}.} PSBT, short for partially signed Bitcoin transactions, is a Bitcoin standard (BIP-174~\cite{orc20}) that enhances the portability of unsigned transactions and enables multiple parties to sign the same transaction easily.
% A PSBT is created with a set of UTXOs to spend and a set of outputs to receive. Then, the information of each UTXO necessary to create a signature will be added. Once the PSBT is prepared, it can be copied to a program capable of signing it. For multi-signature wallets, this signing step can be repeated using different programs on separate PSBT copies. Multiple PSBTs, each containing one or more necessary signatures, will later be combined into a single PSBT. Finally, the fully signed PSBT can be broadcast via networks.

%--------------------------------
\subsection{PSBT Protcol}

\noindent\textbf{PSBT.} PSBT, short for partially signed Bitcoin transactions, is a Bitcoin standard (BIP-174~\cite{bip174}) for Bitcoin transactions designed to separate the construction and signing processes. Formally, a PSBT can be represented as a tuple \( \mathcal{P} = (\mathcal{T}, \mathbf{In}_{\text{PSBT}}, \mathbf{Out}_{\text{PSBT}}) \), where \( \mathcal{T} \) denotes the underlying Bitcoin transaction \( \mathcal{T} = (txid, In, Out, Witness) \), and \( \mathbf{In}_{\text{PSBT}} \) and \( \mathbf{Out}_{\text{PSBT}} \) encapsulate the additional metadata required for partially signing the transaction.

Each input \( I_i \in \mathbf{In} \) within the transaction \( \mathcal{T} \) corresponds to an input in \( \mathbf{In}_{\text{PSBT}} \), augmented with supplementary information such as the full UTXO details \( (\textbf{txid}^{(i)}, \textbf{vout}^{(i)}, \text{amount}^{(i)}, \text{scriptPubKey}^{(i)}) \), the non-witness UTXO if applicable, and any partial signatures \( \sigma^{(i)} \). Mathematically, the PSBT input can be denoted as:
\[
I_i^{\text{PSBT}} = \left( I_i, \text{UTXO}^{(i)}, \sigma^{(i)} \right)
\]
where \( \text{UTXO}^{(i)} \) provides the necessary context for signing, and \( \sigma^{(i)} \) is the partial signature obtained from one or more signers.

Similarly, each output \( O_j \in \mathbf{Out} \) is associated with \( \mathbf{Out}_{\text{PSBT}} \), which may include additional information such as the redeem script or witness script required for spending conditions. This can be expressed as:
\[
O_j^{\text{PSBT}} = \left( O_j, \text{RedeemScript}_j \right)
\]
ensuring that all participants have access to the necessary scripts to validate and sign the transaction.

The PSBT protocol operates through a series of stages:
\begin{packeditemize}

    \item \textbf{Creation}: An initial PSBT is generated, containing the transaction structure \( \mathcal{T} \), the input references \( \mathbf{In}_{\text{PSBT}} \), and the output definitions \( \mathbf{Out}_{\text{PSBT}} \). At this stage, the PSBT includes all the information required to fully construct the transaction, except for the signatures.
    
    \item \textbf{Signature propagation}: The PSBT is shared among the involved multiple signers. Each participant reviews the PSBT, adds their partial signatures \( \sigma^{(i)} \), and updates the PSBT accordingly. This process can be iterated until all signatures are collected.
    
    \item \textbf{Finalization}: Once all required signatures are present, the PSBT is finalized by embedding the complete signature data into the transaction \( \mathcal{T} \), producing the fully signed transaction \( \mathcal{T}' \). \( \mathcal{T}' \) then is serialized and broadcasted to the Bitcoin network for inclusion in a block.

\end{packeditemize}

Formally, the transformation from a partially signed PSBT \( \mathcal{P} \) to a fully signed transaction \( \mathcal{T}' \) can be depicted as:
\[
\mathcal{T}' = \textit{Finalize}(\mathcal{P})
\]
where the \textit{Finalize} function aggregates all partial signatures \( \sigma^{(i)} \) and integrates them into the corresponding inputs of \( \mathcal{T} \), resulting in a valid and broadcast-ready transaction \( \mathcal{T}' \).

 %--------------------------------
\subsection{BRC20 Transfer via PSBT} 
Building upon the standards of BRC20 and PSBT, we formalize the lifecycle of a PSBT-based BRC20 transfer. Let

\[
\Omega = \{\mathsf{p}, \mathsf{op}, \mathsf{tick}, \mathsf{amt}\}
\]
denote the core inscription metadata that specifies the protocol name (\(\mathsf{p = brc\text{-}20}\)), the operation type (\(\mathsf{op}\)), the token name (\(\mathsf{tick}\)), and the transfer amount (\(\mathsf{amt}\)). In particular, for a transfer operation, we might have:
\[
\Omega = \{\mathsf{p}:\!\mathsf{brc\text{-}20},\;\mathsf{op}:\!\mathsf{transfer},\;\mathsf{tick}:\!\mathsf{ordi},\;\mathsf{amt}:1000\}.
\]

\begin{packeditemize}

\item  {\textbf{Seller’s operation.}}
A seller begins by \emph{inscribing} a satoshi with metadata \(\Omega\) that encodes the intention to transfer \(\mathsf{amt} = 1000\) BRC20 \brc\, (e.g., \(\mathsf{tick} = \mathsf{ordi}\)). Formally, we can regard this inscription event as creating a tuple 
\[
\mathcal{I} = \bigl(\textbf{txid}_s, \textbf{vout}_s, \Omega\bigr),
\]
where \(\textbf{txid}_s\) and \(\textbf{vout}_s\) identify the unspent output hosting the inscribed satoshi. The existence of \(\mathcal{I}\) indicates the seller’s willingness to exchange these tokens for a certain Bitcoin amount (e.g., \( 0.2\btc \) \,).

\smallskip
\item  {\textbf{Creation of PSBT.}}  
Next, the seller assembles a partially signed transaction \(\mathcal{P}\) that integrates \(\mathcal{I}\) as one of the inputs. Concretely, the seller designates an output 
\[
{O_s} = (\,0.2\btc~\, ,~\text{scriptPubKey}_s)
\]
to reflect the requested price. In other words, by offering an output crediting \(0.2\btc\) \, to one of the seller’s addresses, the seller stipulates that any buyer must fund this output in exchange for the \(\mathsf{amt} = 1000\) \brc\, inscribed on the satoshi from \(\textbf{txid}_s, \textbf{vout}_s\). The rest of the transaction structure (including any change outputs to the buyer or additional \(\texttt{witness}\) fields referencing \(\Omega\) is specified but not fully signed.

\smallskip
\item  {\textbf{Publishing the PSBT.}}  
The seller then publishes \(\mathcal{P}\) to a marketplace or similar off-chain platform, making it visible to prospective buyers. The marketplace now holds a partially signed transaction 
\[
\mathcal{T} = \mathcal{P}\bigl(\mathbf{In}_{\mathrm{mkt}}, \mathbf{Out}_{\mathrm{mkt}}\bigr),
\]
which includes the vital metadata \(\Omega\) describing the token type and amount. At this stage, no complete signature for the spending input \(\mathcal{I}\) is present, or only a partial signature belonging to the seller exists (depending on multi-sig policies).

\smallskip
\item  {\textbf{Buyer’s operation.}}  
A buyer who wishes to acquire \(1000\) \(\mathsf{ordi}\)\brc\, must \emph{finalize} \(\mathcal{P}\) by providing the necessary \(0.2\,\btc~\, \) input(s) to the transaction and appending their own signatures. Suppose the buyer’s contribution is a UTXO \(\bigl(\textbf{txid}_b, \textbf{vout}_b, \text{amount}_b\bigr)\) such that \(\text{amount}_b \ge 0.2\,\btc~\)~. The buyer modifies 
\[
\mathbf{In}_{\text{PSBT}} \;\cup\; \{(\textbf{txid}_b, \textbf{vout}_b)\}
\]
to incorporate this new input, ensuring the resulting transaction covers the required payment to \(\text{scriptPubKey}_{\mathrm{s}}\). The buyer then attaches any partial signatures \(\sigma_b\) necessary to authorize the spending of \(\bigl(\textbf{txid}_b, \textbf{vout}_b\bigr)\).

\smallskip
\item   {\textbf{Finalizing the PSBT.}}  
When all inputs are properly signed—i.e., the seller’s signature on \((\textbf{txid}_s, \textbf{vout}_s)\) plus the buyer’s signature on \((\textbf{txid}_b, \textbf{vout}_b)\)—the PSBT is transformed into a valid raw transaction. One may express this as
\[
\mathcal{T}' \;=\; \textit{Finalize}\bigl(\mathcal{P}\bigr),
\]
where \(\mathcal{T}'\) denotes the fully signed Bitcoin transaction. Once the buyer broadcasts \(\mathcal{T}'\) to the network, nodes incorporate it into the mempool and, upon successful validation and block inclusion, the transaction becomes irrevocable.

\smallskip
\item {\textbf{Off-chain state updates.}} 
Subsequent to on-chain confirmation, off-chain ledgers or token-management layers update the final token states \(\mathcal{S}\). Let 
\[
\left\{
\begin{aligned}
\mathcal{S}_\mathrm{buyer} &\leftarrow \mathcal{S}_\mathrm{buyer} + 1000\brc~ , \\
\mathcal{S}_\mathrm{seller} &\leftarrow \mathcal{S}_\mathrm{seller} - 1000\brc~ ,
\end{aligned}
\right.
\]
denote the change in \(\mathsf{ordi}\) \brc~ balance, assuming an off-chain indexer or marketplace logic tracks the BRC20 token state. At the same time, the seller’s on-chain balance increases by 0.2\btc~, while the buyer’s on-chain balance decreases accordingly. In other words, the cryptographic settlement on Bitcoin triggers an atomic exchange of the BRC20 tokens for \(\mathrm{BTC}\), ensuring \(\mathcal{T}'\) captures the binding transaction logic. 

\end{packeditemize}

\smallskip
It is worth noting that the protocol necessitates two on-chain transactions to finalize the \textcolor{teal}{$\mathsf{Transfer}$} operation. The first transaction is to signify the action, while the second is an actual transaction on-chain  (detailed mechanism refers to \cite{qi2024brc20}). Our attack focuses on the second transaction.

%================================================= 
\section{BRC20 Snipping Attack}
\label{sec-attack}
%================================================= 
We present a vulnerability tied to the way BRC20 transactions are published using PSBT, termed \textit{BRC20 snipping attack}.

\subsection{Attack Description}
A snipping attack in the context of BRC20 tokens on Bitcoin is a focused form of fee-driven front-running that targets a PSBT before it is fully confirmed. Its defining feature involves intercepting the partially signed data and crafting a competing transaction that replicates the essential transfer details. Unlike standard front-running methods, which typically exploit fee escalations, a snipping attack seizes upon the unique conditions of partially signed workflows. By broadcasting a version of the transaction with a higher fee than the legitimate proposal, the attacker ensures that miners will prioritize the malicious transaction, thus rendering the original transaction invalid or unconfirmable. 

\subsection{Motivation}
The snipping attack typically manifests in a marketplace environment where BRC20 tokens are traded using partially signed transactions. In such a setting, sellers initiate token transfers by creating PSBTs that outline the details of the transaction, including the amount of tokens to be transferred and the associated Bitcoin inputs and outputs. These PSBTs are often shared among multiple participants, such as buyers, sellers, and possibly escrow services, to gather the necessary signatures before the transaction is finalized and broadcasted to the network. This collaborative workflow, while facilitating multi-party authorization, inadvertently exposes the PSBT to potential interception by malicious actors.

Within this context, multiple buyers may concurrently compete to finalize a token transfer, each aiming to have their transaction confirmed promptly to secure the desired tokens. The motivations driving an attacker to execute such an attack are multifaceted.
\begin{packeditemize}
    \item The attacker seeks financial gain by redirecting valuable BRC20 tokens to their own address, stealing assets intended for legitimate buyers.
    \item By invalidating the original transaction, the attacker can cause a denial of service, disrupting the intended transfer and potentially causing economic losses for both buyers and sellers.
    \item The attacker may also aim to manipulate the market by affecting token prices or eroding buyer trust. Repeated successful snipping attacks can lead to a loss of confidence in the marketplace and destabilize the overall BRC20 token economy.
\end{packeditemize}

\subsection{Attack Methodology}
\begin{figure}[!h]
    \centering
    \includegraphics[width=0.9\linewidth]{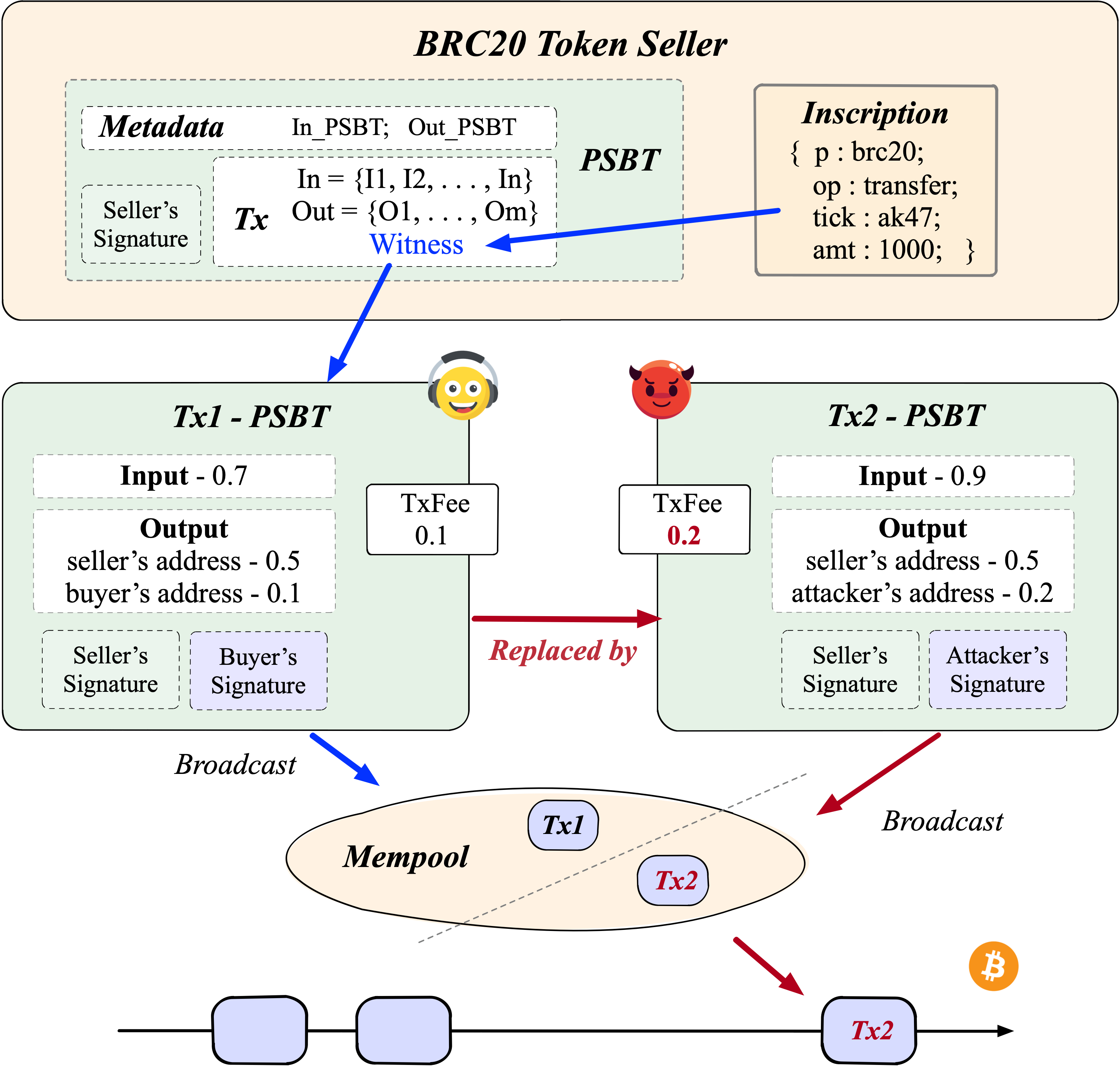}
    \caption{Workflow of Our Snipping Attack}
    \label{fig:workflow}
   % \vspace{-0.15in}
\end{figure}

Fig.\ref{fig:workflow} and the steps below illustrate how an attacker with minimal access can exploit PSBT transactions. The workflow assumes that the attacker can observe or intercept the PSBT metadata and that they possess UTXOs sufficient to cover the total requested output value plus a higher fee.

A crucial element of this strategy is timing. The attacker must insert the malicious transaction into the mempool while the legitimate one remains unconfirmed. The attacker’s high-fee option prompts miners to drop or invalidate the original partial transaction due to conflicting inputs. Additionally, by retaining the core BRC20 inscription data, the attacker ensures that the on-chain record still recognizes the same token transfer parameters, allowing them to channel the tokens to an address of their choosing.

\begin{packeditemize}
\item 
\textbf{Step 1: Intercept the legitimate PSBT.}
By observing unconfirmed BRC20 transactions (often embedding JSON-like data within the witness field) in the mempool, the attacker is able to identify the legitimate buyer’s pending PSBT \(\mathcal{P}\).

\item 
\textbf{Step 2: Parse BRC20 transfer details.}
Extract essential fields from $\mathcal{P}$:
\begin{enumerate}
    \item The spending inputs: $\{( \textbf{txid}^{(i)}, \textbf{vout}^{(i)})\}$;
    \item The receiving address(es) and amounts for the BRC20 context, e.g., $\mathsf{p}: \mathsf{brc20}, \mathsf{op}: \mathsf{transfer}, \mathsf{tick}, \mathsf{amt}, \dots$;
    \item The on-chain fee estimate $f_{\mathrm{legit}}$ if deducible from the input-output difference.
\end{enumerate}

\item\textbf{Step 3: Construct the malicious transaction.}
Using attacker-owned UTXOs $\mathcal{A}$, prepare a new PSBT $\mathcal{P}_{\mathrm{atk}}$ structure referencing equivalent BRC20 metadata. Let
\[
\mathcal{P}_{\mathrm{atk}} = ( \mathbf{In}_{\mathrm{atk}}, \mathbf{Out}_{\mathrm{atk}}, \mathbf{Witness}_{\mathrm{atk}} ).
\]
\begin{enumerate}
    \item Copy the core inscription data (e.g., \textit{tick, amt, op}) to maintain validity for the same BRC20 token;
    \item Allocate outputs so that at least one matches the seller’s address and amount (if the aim is to mimic the buyer’s payment);
    \item Reserve minimal change for the attacker’s address, ensuring a higher fee $f_{\mathrm{atk}} > f_{\mathrm{legit}}$.
\end{enumerate}

\item\textbf{Step 4: Sign and finalize.}
Use the attacker’s private key(s) to sign $\mathcal{P}_{\mathrm{atk}}$ fully. Ensure the transaction meets all constraints (e.g., scriptPubKey validations, input availability). Once signed,
\[
\mathcal{T}_{\mathrm{atk}} \; \leftarrow \; \mathrm{finalizePSBT}(\mathcal{P}_{\mathrm{atk}}).
\]

\item\textbf{Step 5: Broadcast malicious transaction.}
Send $\mathcal{T}_{\mathrm{atk}}$ to the mempool before the legitimate PSBT is confirmed. If the fee surpasses the original transaction’s fee, miners will likely prioritize $\mathcal{T}_{\mathrm{atk}}$ over the legitimate one.

\item\textbf{Step 6: Observe mempool and block inclusion.}
Monitor the Bitcoin mempool for confirmation. If $\mathcal{T}_{\mathrm{atk}}$ is included in the next block, the legitimate transaction becomes invalid due to input conflicts, causing it to be removed from the mempool.

\end{packeditemize}

%================================================= 
\section{Local Experiment}
\label{sec-imple}
%================================================= 
%We established a controlled local experiment simulating a BRC20 snipping attack. 

\subsection{Experimental Setup}

\noindent\textbf{Hardware setup.}
The experiment was conducted on a desktop computer equipped with an Intel Core i7-9700K processor running at 3.60 GHz, 32 GB of DDR4 RAM, and a 1 TB NVMe SSD. The operating system used was Ubuntu 20.04 LTS.

\smallskip
\noindent\textbf{Software setup.} The core of our experimental setup relied on Bitcoin Core version 26.0, configured to operate in Regtest mode. Regtest provides a private Bitcoin environment that allows for rapid block generation and fine-grained control over network conditions, making it ideal for testing purposes. Specifically, we initiated a Bitcoin Core node, configured with the following parameters: \textit{-regtest} to enable Regtest mode and \textit{-txindex=1} to maintain a complete transaction index.

We utilized Bitcoin Core's built-in PSBT functionalities (e.g., \texttt{createpsbt}, \texttt{finalizepsbt}, etc.) to manage the creation and signing of BRC20 token transactions. Furthermore, we implemented the BRC20 protocol by building upon the existing open-source Oridinal~\cite{ordinalwallet} project. To achieve accurate and real-time tracking of BRC20 token states, we deployed locally running indexing services (writing in Python scripts) that parse and monitor the JSON-based inscriptions embedded within transaction \textit{witness} fields.

\subsection{Threat Model}
In our local experiment, We assume the attacker can monitor or obtain transaction data in the following ways:
\begin{itemize}
    \item \textbf{Access to PSBT:} The attacker is able to observe or intercept PSBTs before they are fully signed and broadcast. Such interception could occur through compromised communication channels (e.g., local file sharing). 
    \item \textbf{Mempool visibility:} For already completed PSBT transactions, the attacker can inspect the local mempool. This visibility allows the attacker to detect any transaction referencing the same UTXOs. In a real-world scenario, similar transparency might arise from a public Bitcoin data explorer (i.e., mempool.space~\cite{mempool2024}).
\end{itemize}

Additionally, we also assume that the attacker controls at least one UTXO with enough bitcoin to create a higher-fee transaction than legitimate buyers. If the attacker needs to complete a partial signature, they can also generate the necessary private keys for their own transaction inputs.

\subsection{Attack Strategy}

\smallskip
\noindent\textbf{Attack goal.} The attacker’s primary goal in our local experiment is to hijack an in-progress BRC20 token transfer by exploiting vulnerabilities in the PSBT workflow. In other words, the attacker invalidates or preempts the original transaction, "snipping" the intended BRC20 token transfer.

% Specifically, the attacker seeks to "snip" the legitimate PSBT, modify its fee, and re-broadcast the manipulated version so that it is confirmed first within the local blockchain environment. By achieving this, the attacker invalidates or preempts the original transaction, "snipping" the intended token transfer. 

\smallskip
\noindent\textbf{Attack method.} Shown in Algorithm~ \ref{alg:brc20_snipping_attack}, to execute this attack, the attacker first obtains the PSBT by monitoring mempool. Because the local node logs incomplete PSBT data or temporarily loads it into memory while awaiting additional signatures, the attacker can parse those logs or memory dumps. Next, they modify the transaction metadata (e.g., UTXO information and output address), most critically, increasing the miner fee (i.e., decreasing the change sent back to the sender) beyond what the legitimate sender has offered. The attacker then completes partial signatures to render the malicious transaction valid. Finally, they re-broadcast this high-fee variant to the local mempool, where the node’s incentive mechanism causes mining simulators in the regtest environment to prioritize it over the legitimate transaction. 

\begin{algorithm}[t]
\caption{BRC20 Snipping Attack Methodology}

\renewcommand{\arraystretch}{1.30}
\footnotesize
\setlength{\tabcolsep}{4pt}

\textbf{Global Parameters:} \\
\(\mathcal{I}_\text{buyer}\), \(\mathcal{I}_\text{atk}\), Attacker's fee rate \(f_\text{atk}\), Buyer's fee rate \(f_\text{buyer}\), Mempool \(C_m\).

\vspace{0.1cm}

\begin{algorithmic}[1]
\State \textbf{Step-\ding{172}: Attacker monitors the mempool.}
\State Attacker $\mathcal{A}$ monitors mempool for unconfirmed BRC20 transactions.
    \State \quad Attacker observes PSBT $\mathcal{P}$ with input \(\mathcal{I}_\text{buyer} = (\textbf{txid}_b, \textbf{vout}_b)\) and output \(\mathcal{O}_s = (0.5\btc~\,, \text{scriptPubKey}_s)\).
    \State Attacker identifies fee rate \(f_\text{buyer}\) associated with the transaction.

\noindent\makebox[\linewidth]{\rule{0.99\linewidth}{0.3pt}}

\State \textbf{Step-\ding{173}: Attacker creates sniping PSBT.}
\State Attacker $\mathcal{A}$ prepares sniping transaction $\mathcal{P}_\text{atk}$ targeting \(\mathcal{I}_\text{buyer}\).
    \State \quad Attacker uses their own UTXO \(\mathcal{I}_\text{atk} = (\textbf{txid}_a, \textbf{vout}_a)\) as input to match \(\mathcal{I}_\text{buyer}\).
    \State \quad Attacker sets a higher fee \(f_\text{atk}=0.28125\btc ~> f_\text{buyer}=0.27985\btc~\,\).
    \State Create sniping output \(O_\text{atk} = (0.5\btc~\,, \text{scriptPubKey}_s)\), ensuring the payment goes to the seller’s address.
    \State Add additional change output to attacker’s address \(\mathcal{O}_\text{change} = (0.00005\btc~\,, A_\mathcal{A})\).
    \State Broadcast \(\mathcal{P}_\text{atk}\) to mempool.

\noindent\makebox[\linewidth]{\rule{0.99\linewidth}{0.3pt}}

\State \textbf{Step-\ding{174}: Mempool conflict resolution.}
\State Mempool processes \(\mathcal{P}_\text{atk}\) and compares fee rates.
    \If{Attacker's fee \(f_\text{atk} > f_\text{buyer}\)}
        \State \quad $\mathcal{T}_\text{atk}$ is prioritized for inclusion in the next block.
        \State \quad \textit{Legitimate transaction \(\mathcal{T}_\text{buyer}\) is invalidated due to input conflict and removed from mempool.}
    \Else
        \State \quad \textit{Transaction not included due to low fee.}
    \EndIf

\noindent\makebox[\linewidth]{\rule{0.99\linewidth}{0.3pt}}

\State \textbf{Step-\ding{175}: Transaction Confirmation.}
\State Confirm the inclusion of \(\mathcal{T}_\text{atk}\) in the next block.
    \State \quad Verify that \(\mathcal{T}_\text{atk}\) is confirmed, while the original \(\mathcal{T}_\text{buyer}\) remains unconfirmed or discarded.

\noindent\makebox[\linewidth]{\rule{0.99\linewidth}{0.3pt}}

\State \textbf{Step-\ding{176}: Verify Attack Success.}
\State Attacker's wallet (\(A_{\mathcal{A}}\)) should now hold the BRC20 tokens transferred through the sniping attack.
    \State \quad Confirm the amount of \(\mathsf{ak47}\) tokens in \(A_{\mathcal{A}}\) matches the amount originally inscribed in \(\mathcal{I}_\text{buyer}\).
    \State \quad Confirm that \(\mathcal{T}_\text{buyer}\) is no longer valid in the mempool.

\end{algorithmic}
\label{alg:brc20_snipping_attack}
\end{algorithm}

\smallskip
\noindent\textbf{Success Criteria.}
We deem the snipping attack successful if the attacker’s malicious transaction both confirms before (and thereby invalidates) the legitimate transaction \emph{and} delivers the targeted BRC20 tokens to the attacker’s control. 

Concretely, two conditions must be met:

\begin{packeditemize}
    \item \textit{Block inclusion.} The manipulated transaction authored by the attacker is accepted in a newly mined block prior to the genuine transaction, causing the genuine transaction to remain unconfirmed or be evicted from the mempool due to input conflicts.

    \item \textit{Token receipt.} Upon confirmation of the attacker’s transaction, a query to local indexers indicates that the BRC20 tokens, originally intended for the legitimate buyer, have been credited to an address controlled by the attacker. Verifying an increase in the attacker’s token balance confirms the successful appropriation of the asset.
\end{packeditemize}

\subsection{Execution of the Attack}

\smallskip
\noindent\textbf{Seller address setup.}
To initiate our experiment, we first created a seller wallet and an address within the \textit{bitcoind-regtest} network, specifically \textcolor{teal}{$\text{adr.0dd8}$}. We then used the built-in mining functions in regtest mode to generate blocks, designating the newly created address to receive the corresponding block rewards. After the mining stage, a balance check on the seller’s wallet confirmed that it held exactly 1.5625\btc~, which would serve as the principal for creating and issuing BRC20 tokens, as well as covering transaction fees.

\smallskip
\noindent\textbf{Token deploy and mint.}
We used the same seller address to mint a BRC20 token on the regtest network. Following the BRC20 protocol, we prepared a JSON-based inscription containing the necessary parameters: \{$\mathsf{p:brc20}$; $\mathsf{op:deploy}$;  $\mathsf{tick:ak47}$;
 $\mathsf{max:2100000}$; $\mathsf{lim:1000}$\}.
This metadata was then serialized into hexadecimal form, yielding \textcolor{teal}{hex.307d}. 

Subsequently, we constructed a raw transaction on our regtest network using the \texttt{createrawtransaction} command, consuming the output from a prior transaction (\textcolor{teal}{txid.240c}) as input, and specifying the hexadecimal-encoded inscription (\textcolor{teal}{hex.307d}) as \textit{"data"} in the transaction output. After finalizing and signing the transaction with our seller wallet, we broadcast it to the local network and mined a new block to confirm it. The transaction was successfully included in \textcolor{teal}{block.a2e2}, marking the on-chain deployment of the BRC20 \textit{ak47} \brc.

 In our local regtest environment, BRC20 tokens default to belonging to the deployer’s address once the “deploy” transaction confirms. Consequently, the seller’s wallet automatically acquired a total of \(\textit{21,000,000}\) \(\textit{ak47}\) \brc\, without requiring a separate \texttt{mint} operation. 

\smallskip
\noindent\textbf{Token transfer via PSBT.}
To initiate a subsequent transfer, the seller prepared a JSON inscription reflecting the \texttt{transfer} operation: \{$\mathsf{p:brc20}$; $\mathsf{op:transfer}$;  $\mathsf{tick:ak47}$; $\mathsf{amt:1000}$\}. Converting this metadata to a hexadecimal string yielded: \textcolor{teal}{hex.227d}.

Using the \(\texttt{createpsbt}\) function, the seller constructed a new partially signed transaction. The relevant input referenced a previous transaction output (\(\texttt{txid:}\) \textcolor{teal}{tx.641d}, \(\texttt{vout: 0}\)) held by the seller, while the output was defined to include the transfer inscription in its \(\texttt{"data"}\) field. Executing \(\texttt{createpsbt}\) produced the following PSBT string: \textcolor{teal}{psbt.r85P}.

Next, we used \(\texttt{walletprocesspsbt}\) to sign this PSBT. However, as only one signature was provided and the PSBT did not yet meet all necessary signing requirements for a complete transaction, the \(\texttt{walletprocesspsbt}\) result indicated \(\texttt{"complete": false}\). This status signifies that, under our experimental setup, additional signatures would be needed before broadcasting the final transaction.

\smallskip
\noindent\textbf{Buyer \& attacker addresses setup.}
We introduced three additional addresses to represent potential buyers and attackers. The ordinary buyer’s address was set to \textcolor{teal}{adr.qa8r}, while two attacker addresses, \textcolor{teal}{adr.v6r9} and \textcolor{teal}{adr.xffg}, were designated to simulate different fee-bidding strategies. Specifically, one attacker was intended to surpass the ordinary buyer’s fee, while the other remained lower, helping us observe various outcomes in our snipping experiments.

\smallskip
\noindent\textbf{Buyer signs PSBT.}
First, we assumed that the ordinary buyer discovered the seller’s partially signed BRC20 transaction. From the buyer’s perspective, the crucial element of interest was the transaction’s \texttt{data} field, which encodes the BRC20 transfer details in hexadecimal format. Using these details, the buyer constructed a new and fully formed PSBT with the following command (simplified for illustration):

\begin{lstlisting}[language=bash, caption={Constructing a new PSBT from buyer’s view}]
bitcoin-cli -regtest createpsbt
'[{"txid": "6ec491e4928d9b6af81b0ebd5040e1a44834f
2576e1085ecb79d5fddc4fbb2ea", "vout": 0}]'
'[
  {"bcrt1qkckuparv0d8nxadvzxgl0m0fmsn0ym6xfq0dd8": 0.5},
  {"bcrt1qackm4gsk5szqmylr0k5fq2n6v9jugz4473qa8r": 0.1},
  {
    "data": "7b22703a226272632d3230222c226f70223
    a227472616e73666572222c227469636b223a22616b34
    37222c22616d74223a2231303030227d"
  }
]'
\end{lstlisting}

In this command:
\begin{itemize}
    \item \textcolor{teal}{tx.b2ea} (\texttt{vout=0}) is the UTXO the buyer intends to spend (with the amount of 0.87985\btc~).
    \item \textcolor{teal}{adr.0dd8} (the seller’s address) is allocated 0.5\btc\,.
    \item \textcolor{teal}{adr.qa8r} (the buyer’s address) is allocated 0.1\btc\,.
    \item The \texttt{data} field includes the hexadecimal-encoded BRC20 transfer instruction, representing the inscription \{$\mathsf{p:brc20}$; $\mathsf{op:transfer}$;  $\mathsf{tick:ak47}$;  $\mathsf{amt:1000}$\}.
\end{itemize}

The input UTXO provides 0.87985\btc~in total, while the outputs allocate 0.5\btc~and 0.1\btc~, respectively. According to Formula~\ref{tx-fee}, the resulting fee $\mathcal{F}$ is:
   $$
            \mathcal{F}_b= 0.87985\btc~ - (0.5\btc~ + 0.1\btc~\,) = 0.27985\btc~\,.
    $$

After creating this new PSBT, the bitcoin-cli command returned a base64-encoded string (\textcolor{teal}{psbt.+Gqb}) indicating the partially constructed transaction. The buyer used the \texttt{walletprocesspsbt} command to sign this PSBT with their private key, returning hex-encoded transaction data (\textcolor{teal}{hex.e140}). A response looks like:

\begin{lstlisting}[language=bash,caption={Buyer-signed PSBT Example}]
{
"psbt": "cHNidP8BALQCAAAAAeqy+8TdX5237IUQblfyNEik4UBQvQ4b+GqbjZLkkcRuAAAAAAD9////A4Dw+gIAAAAAFgAUti3A9Gx7TzN1rBGR9+3p3Cbyb0bgIgIAAAAAABYAFOZzVbqGcZy..."
"complete": true
"hex": "02000000000101eab2fbc4dd5f9db7ec85106e57f
23448a4e140...8200000000"
}
\end{lstlisting}

\texttt{"complete": true} now indicates the transaction has all required signatures for this scenario, meaning it can be broadcast and confirmed without additional inputs.

%\begin{figure}[h]
%    \centering
%    \includegraphics[width=\linewidth]{img/psbt-normal.png}
%    \caption{Details of Buyer-signed PSBT}
%    \label{fig:tx-fee}
%    \vspace{-0.15in}
%\end{figure}

\smallskip
\noindent\textbf{Broadcast the completed PSBT.}
The final step is to broadcast the fully signed transaction to the local regtest network. Having obtained the hex-encoded transaction data (\textcolor{teal}{hex.e140}) from the completed PSBT, the buyer invokes the \texttt{sendrawtransaction} command, supplying the raw transaction in hexadecimal form and returning a transaction id \textcolor{teal}{txid.c936}. This submission passes the transaction into the local node’s mempool for validation. Once validated, the node broadcasts the transaction to the rest of the regtest environment, enabling block producers to mine it. Under non-adversarial circumstances, this broadcasted transaction would then be included in a newly mined block, completing the token transfer from the seller to the buyer. However, in our subsequent snipping attack experiments, the attacker will attempt to outbid this transaction, thereby invalidating the transfer.

\smallskip
\noindent\textbf{Adverisal buyers (attackers) snip PSBT.}
In our experimental setup, the attacker aims to invalidate the buyer’s PSBT by crafting and broadcasting an alternative PSBT with a higher fee. Two attacker addresses were introduced to conduct a comparative experiment: one attempting to outbid the ordinary buyer’s fee, and the other offering a lower fee to illustrate a failed snipping attempt. 

\smallskip
\noindent\textit{(1) \underline{Snipping with a higher fee}.}
The attacker’s address (\textcolor{teal}{$\text{adr.v6r9}$}) had a UTXO valued at 0.7813\btc\,. This UTXO provides enough funds to both pay the seller the same amount (0.5\btc~) that the buyer intended to pay, and simultaneously offer a fee slightly exceeding the legitimate transaction fee of 0.27985\btc\,. By allocating only 0.00005\btc~back to the attacker’s own address, the resulting fee is calculated to be 0.28125\btc\,, ensuring that this snipping transaction becomes more appealing to local miners or block producers on regtest.
   $$
            \mathcal{F}_h= 0.7813\btc~ - (0.5\btc~ + 0.00005\btc~\,) = 0.28125\btc~\,.
    $$

The attacker mirrors the buyer’s steps of using \(\texttt{createpsbt}\) but replaces the inputs and outputs with their own addresses and amounts. In particular, they preserve the essential details that must remain consistent—namely sending 0.5\btc~to the seller’s address (\textcolor{teal}{$\text{adr.0dd8}$}) and embedding the same BRC20 transfer inscription in the \(\texttt{data}\) field. An invocation is shown below:

\begin{lstlisting}[language=bash, caption={Constructing the Attacker’s PSBT with a Higher Fee}]
bitcoin-cli -regtest createpsbt 
'[{"txid": "6ec491e4928d9b6af81b0ebd5040e1a44834f2576
e1085ecb79d5fddc4fbb2ea", "vout": 0}]'
'[
  {"bcrt1qkckuparv0d8nxadvzxgl0m0fmsn0ym6xfq0dd8": 0.5},
  {"bcrt1qtpq478leuks3wcg9zgr7rqhz6enszpsmt7v6r9": 0.00005},
  {
    "data": "7b22703a226272632d3230222c226f70223a
    227472616e73666572222c227469636b223a22616b343
    7222c22616d74223a2231303030227d"
  }
]'
\end{lstlisting}

This step yields a base64-encoded PSBT string (\textcolor{teal}{psbt.UBQv}). The attacker proceeds to sign this PSBT (through \(\texttt{walletprocesspsbt}\)). A successful signature produces a valid, fully signed transaction in hexadecimal form (\textcolor{teal}{hex.ec85}).

%\begin{figure}[h]
%    \centering
%    \includegraphics[width=\linewidth]{img/psbt-high.png}
%    \caption{Scenario 1: Attacker-signed PSBT with Higher TxFee)}
%    \label{fig:tx-high}
%    \vspace{-0.15in}
%\end{figure}

Finally, the attacker issues \(\texttt{sendrawtransaction}\) to broadcast this signed raw transaction to the local regtest mempool. Upon success, the node returns a \(\texttt{txid}\) \textcolor{teal}{txid.f0bb}.

\smallskip
\noindent\textit{(2) \underline{Snipping with a lower fee}.}
In a parallel experiment, we tested a scenario in which the attacker attempts to front-run the buyer’s transaction but uses a notably lower fee than the buyer. This second attacker address, here referred to as \textcolor{teal}{$\text{adr.xffg}$}, holds a UTXO worth 0.700001\btc\,. Following the same basic approach, the attacker preserves the essential fields from the original transfer transaction—namely sending 0.5\btc\, to the seller’s address \textcolor{teal}{$\text{adr.0dd8}$} and embedding the identical BRC20 transfer inscription in the \(\texttt{data}\) field. However, they allocate 0.2\btc~as change to their own address \textcolor{teal}{$\text{adr.xffg}$}, leaving a transaction fee of only 0.000001\btc\,. This amount is far below the ordinary buyer’s fee.
   $$
   \mathcal{F}_l= 0.700001\btc~ - (0.5\btc~ + 0.2\btc~\,) = 0.000001\btc~\,.
    $$

The attacker constructs this low-fee transaction by issuing:
\begin{lstlisting}[language=bash, caption={Constructing the Attacker’s PSBT with a Lower Fee}]
bitcoin-cli -regtest createpsbt 
'[{"txid": "6ec491e4928d9b6af81b0ebd5040e1a44834f
2576e1085ecb79d5fddc4fbb2ea", "vout": 0}]' 
'[
  {"bcrt1qkckuparv0d8nxadvzxgl0m0fmsn0ym6xfq0dd8": 0.5},
  {"bcrt1quee4tw5xwxw236yuw8zyuw83r36dtnztsuxffg": 0.2},
  {
    "data": "7b22703a226272632d3230222c226f7022
3a227472616e73666572222c227469636b223a22616b3437
222c22616d74223a2231303030227d"
  }
]'
\end{lstlisting}

Subsequently, the attacker signs the returned PSBT to generate a fully signed transaction in hexadecimal form and broadcasts this signed transaction, returning a transaction ID (\textcolor{teal}{txid.2323}).

%\begin{figure}[h]
%    \centering
%    \includegraphics[width=\linewidth]{img/psbt-low.png}
%    \caption{Scenario 2: Attacker-signed PSBT with Lower Tx Fee}
%    \label{fig:tx-low}
%    \vspace{-0.15in}
%\end{figure}

\begin{figure}[!htbp]
    \centering
    \subfigure[\textbf{Details of Buyer-signed PSBT:} The Base64-encoded string of the buyer’s signed PSBT (\textcolor{teal}{psbt.+Gqb}) contains the necessary data to complete the transaction. The  highlighted transaction fee shows the calculated fee of \textbf{0.27985000}\btc\,.]{\label{psbt-normal.png}
        \includegraphics[width=\linewidth]{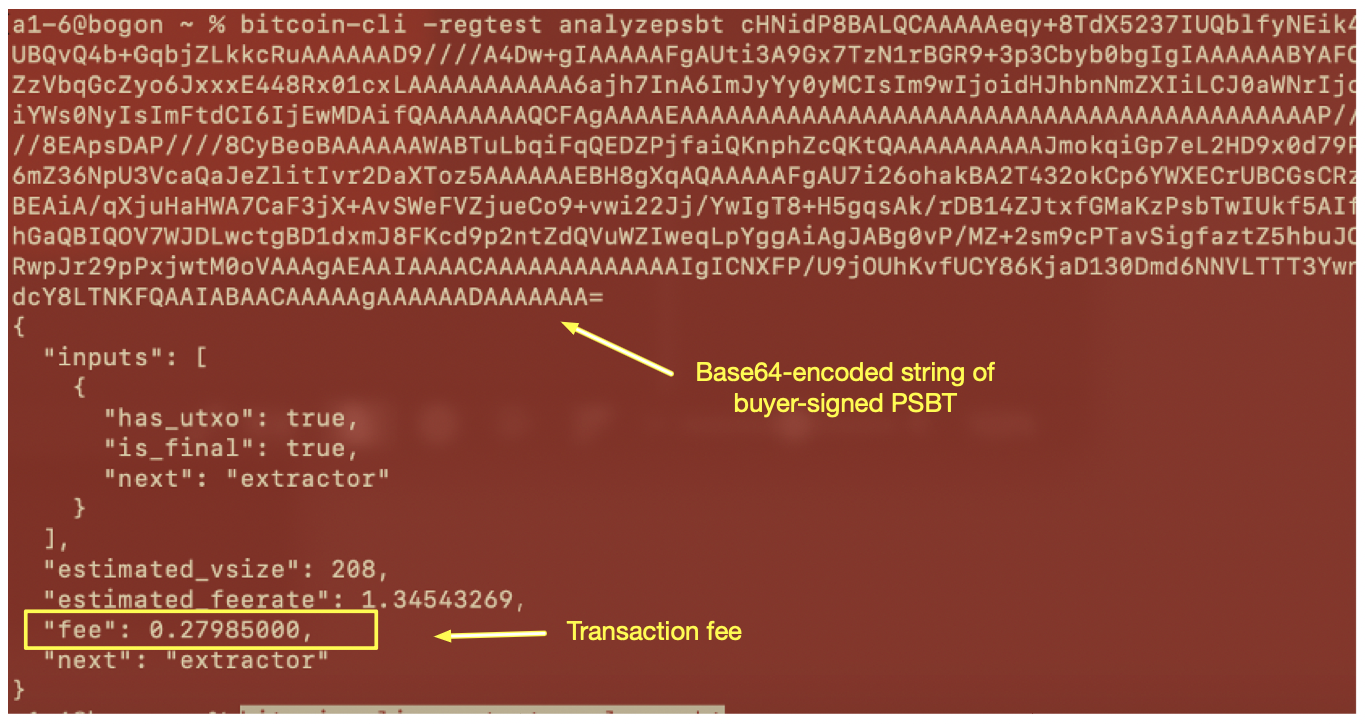}
    }
    \vspace{-0.1in}
    \subfigure[\textbf{Scenario 1: Attacker-signed PSBT with higher TxFee:} It displays the Base64-encoded string of the attacker-signed PSBT (\textcolor{teal}{psbt.UBQv}) with a higher transaction fee. In this scenario, the transaction fee is marked as \textbf{0.28125000}\btc\,, which is higher compared to \ref{psbt-normal.png}, indicating a malicious modification by the attacker.]{\label{fig:tx-high}
        \includegraphics[width=\linewidth]{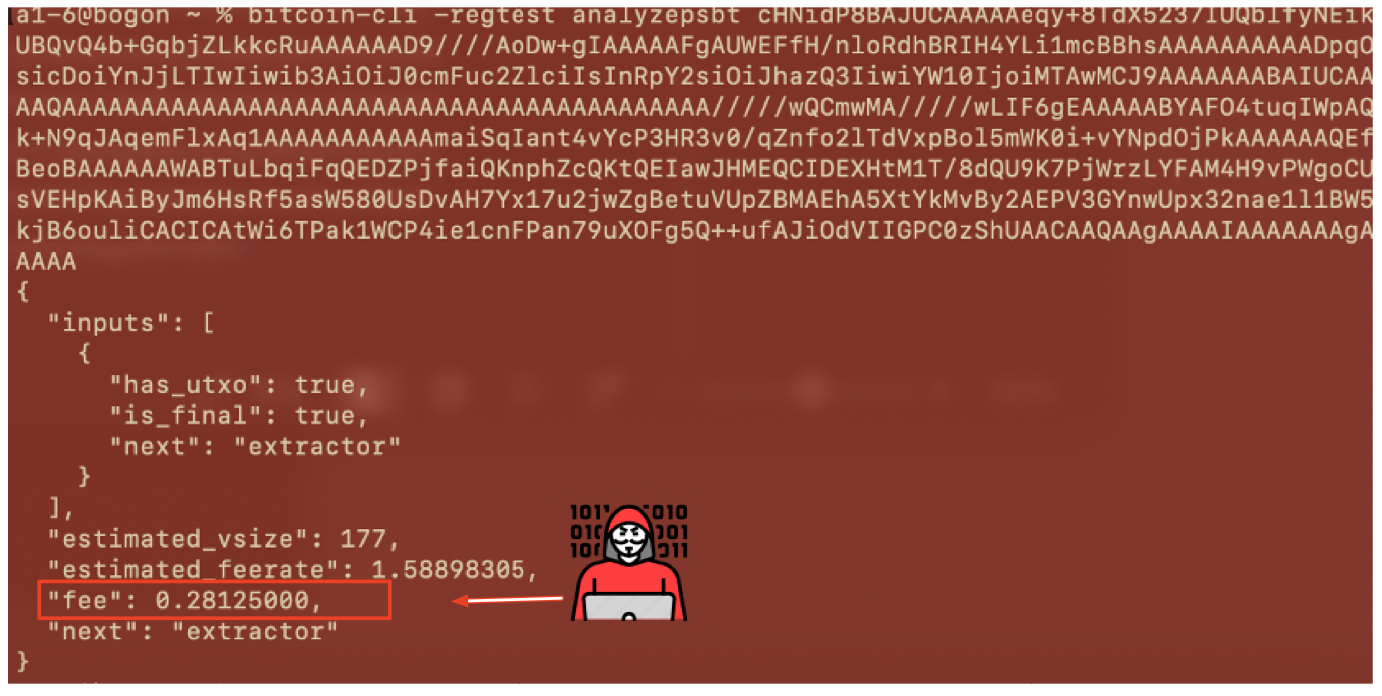}
    }
    \vspace{-0.1in}
    \subfigure[\textbf{Scenario 2: Attacker-signed PSBT with lower TxFee:} It shows the Base64-encoded string of the attacker-signed PSBT (\textcolor{teal}{psbt.b+Gq}) with a lower transaction fee. In this scenario, the transaction fee is significantly reduced to \textbf{0.0000010}\btc\,, which indicates a malicious attempt as a comparison with \ref{fig:tx-high}.]{\label{fig:tx-low}
        \includegraphics[width=\linewidth]{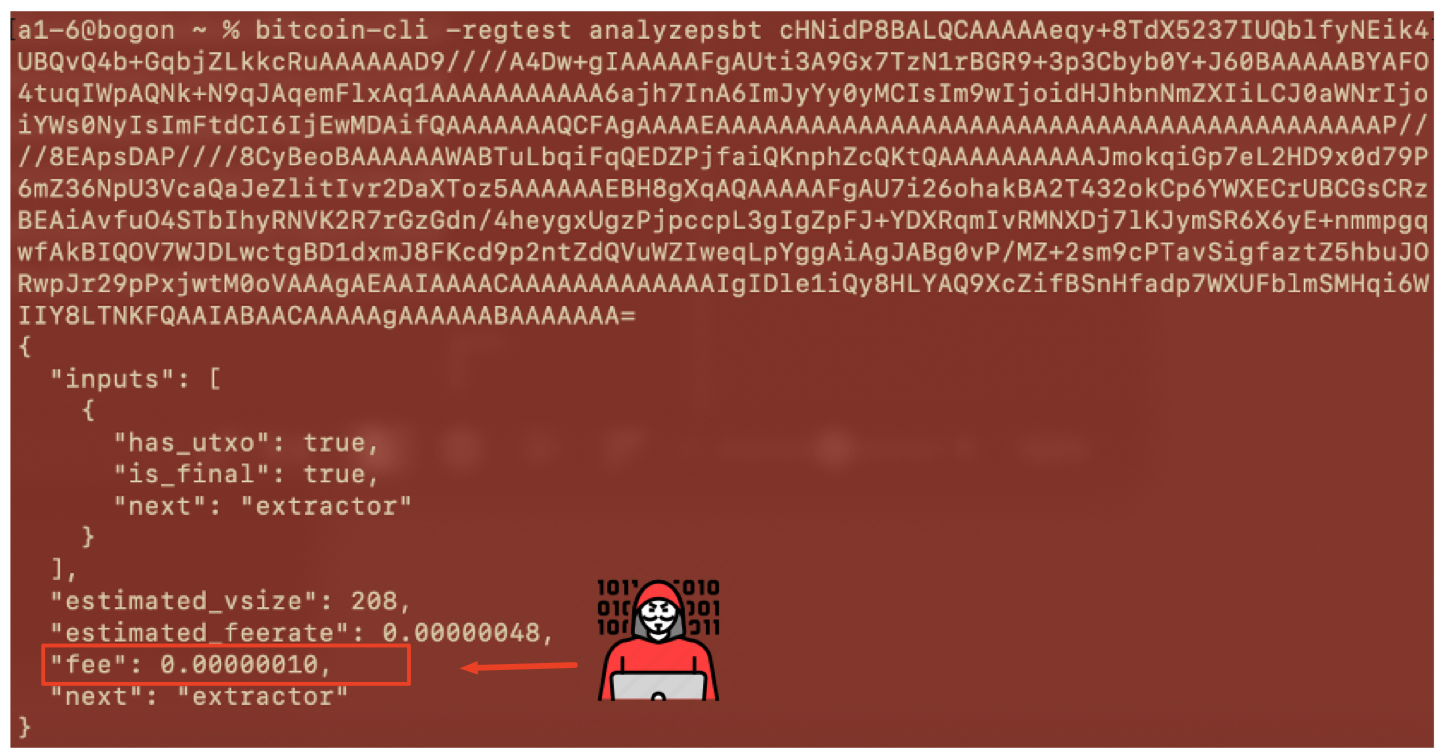}
    }
    \vspace{-0.1in}
    \subfigure[\textbf{Block details after our attack:} There is only one confirmation transaction in the packed block. The transaction is highlighted in the \texttt{tx} field, which is the one (\textcolor{teal}{txid.f0bb}) with the higher transaction fee. This confirms that the attacker’s attack was effective.]{\label{fig:block}
        \includegraphics[width=\linewidth]{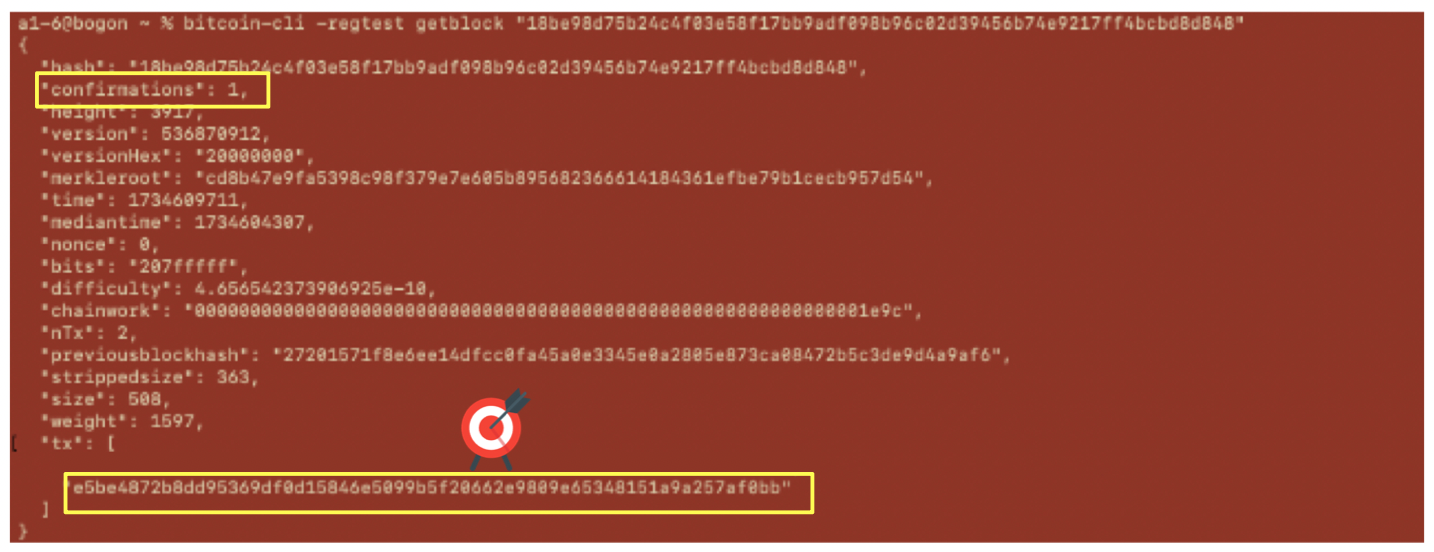}
    }
    \vspace{-0.1in}
    \caption{Screenshots during our attack}
    \vspace{-0.15in}
\end{figure}

\smallskip
\noindent\textbf{Mempool observation.}
Following the attacker’s snipping attempts, we inspected the local mempool using the command \texttt{bitcoin-cli -regtest getrawmempool}. As expected, we discovered three unconfirmed transactions residing in the mempool: the legitimate PSBT transaction broadcast by the ordinary buyer, and the two malicious PSBT transactions issued by the attacker.

Specifically, the mempool listing presents the chronological order of transaction submission. The Ordinary buyer’s transaction (\textcolor{teal}{txid.c936}) is encountered first, immediately followed by the high-fee attacker transaction (\textcolor{teal}{txid.f0bb}), and culminating with the low-fee attacker transaction (\textcolor{teal}{txid.2323}).

\smallskip
\noindent\textbf{Validation of the snipping attack.}
To ascertain whether the attacker’s snipping attempt succeeded, we proceeded to mine a new block in our local regtest environment by issuing:

\begin{lstlisting}[language=bash]
bitcoin-cli -regtest generatetoaddress bcrt1qt30mvqww2cq9nvq4zlefj0l330y4k2ulqtdyks
\end{lstlisting}

This command instructs the regtest node to generate one block and direct its mining reward to the specified address (\textcolor{teal}{adr.tdks}). Upon successful block creation, the node returned a newly mined block hash \textcolor{teal}{block.d848}. Next, we inspected the block’s contents with the returned block hash:
\begin{lstlisting}[language=bash]
bitcoin-cli -regtest getblock 
18be98d75b24c4f03e58f17bb9adf098b96c02d39456b
74e9217ff4bcbd8d848
\end{lstlisting}

\smallskip
\noindent\textit{(1) \underline{First condition - block inclusion.}} In the resulting JSON output (Fig.\ref{fig:block}), the \texttt{tx} field revealed a single relevant transaction ID: \textbf{\textcolor{teal}{tx.f0bb}} which matched the higher-fee PSBT created by the attacker.

To confirm this observation, we rechecked the mempool status via \texttt{getrawmempool} command. At this stage, the previously visible transactions, namely the ordinary buyer’s PSBT and the low-fee attacker transaction, had disappeared from the mempool. This disappearance occurred because those transactions became invalid once the attacker’s higher-fee PSBT transaction was included in the newly mined block. In other words, by consuming the same input PSBT at a higher fee, the attacker’s higher-fee PSBT transaction superseded the other two, causing them to be dropped from the mempool due to a conflict.

\smallskip
\noindent\textit{(2)  \underline{Second condition - token receipt.}} In addition to the confirmation of the attacker’s transaction in the block, we also validate the success of the attack by examining the distribution of BRC20 tokens. Specifically, we check whether the attacker’s wallet has received the expected \texttt{ak47}\brc, confirming that the attack was successful in redirecting the tokens from the legitimate buyer to the attacker. 

To accomplish this, we queried the BRC20 token balances of the attacker’s wallet address (\textcolor{teal}{tx.f0bb}) using a custom script that we developed for tracking token balances across different addresses. The script extracts transaction data from the regtest network and calculates the balance of each address by aggregating the token transfers. Upon performing this check, we observed, as anticipated, that the attacker’s address had received an additional 1000 AK47\brc\,. Furthermore, we verified the token balance of the seller’s address (\textcolor{teal}{tx.0dd8}), which showed a corresponding decrease of 1000 AK47\brc\,.

\begin{figure}[h]
    \centering
    \includegraphics[width=\linewidth]{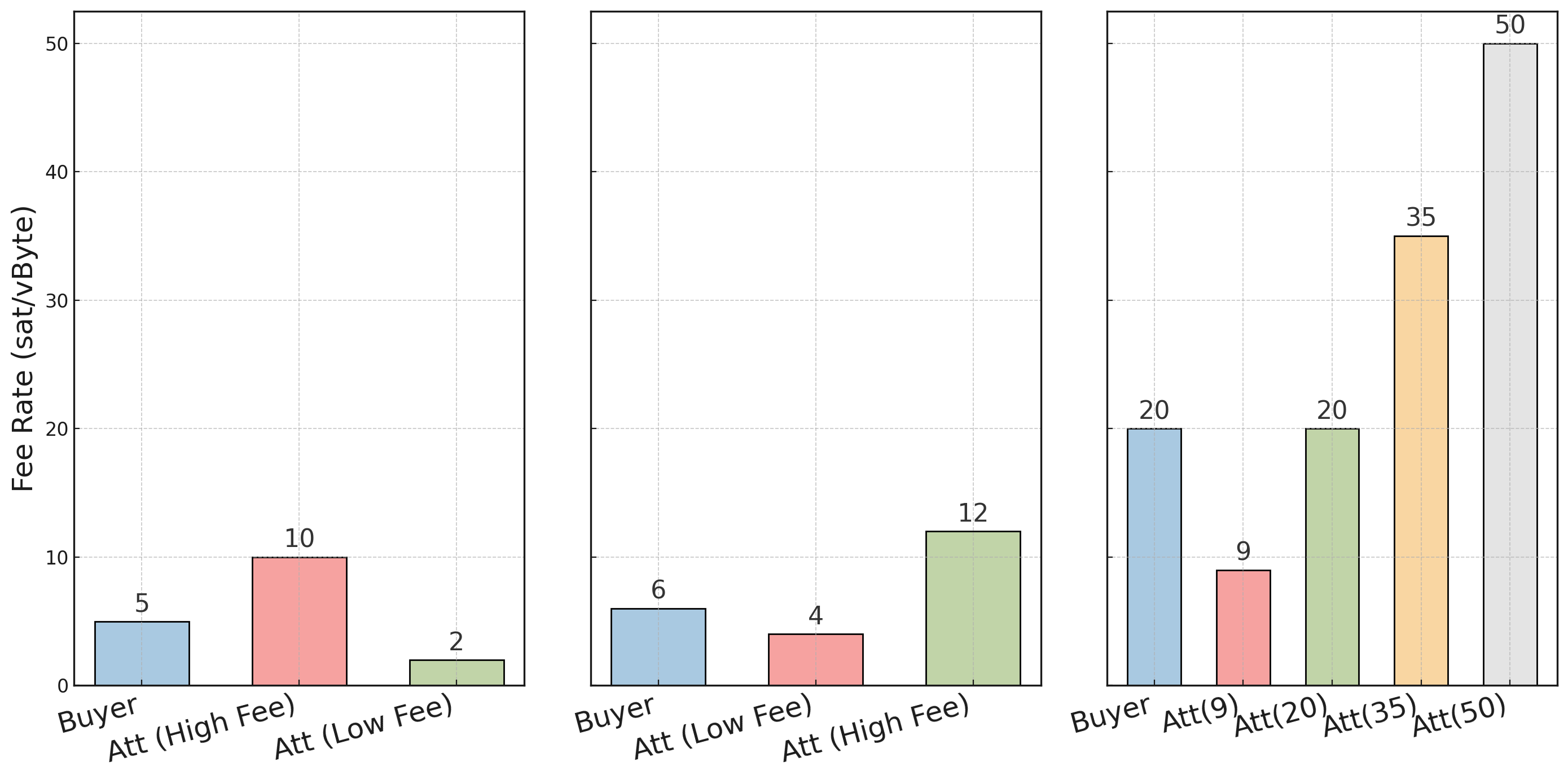}
    \caption{Fee rate comparison in three experiments: The chart compares the fee rates of transactions across three experimental rounds, showing the fee rates of legitimate buyer transactions and those of attackers using high and low fees.}
    \label{fig:comparison}
    \vspace{-0.15in}
\end{figure}

\smallskip
\noindent\textbf{Reproducibility of attack results.}
To ensure the reliability and reproducibility of our findings, we conducted two additional rounds of experiments beyond the initial trial (shown in Fig.~\ref{fig:comparison}). The objective was to further validate whether the observed results were consistent and not due to chance.

In the second round of experiments, we reversed the order of the low-fee and high-fee attacker transactions. This was done to examine whether the sequence in which the transactions were submitted had any impact on the attack's success. Despite the change in order, the results were consistent with the initial experiment: the transaction with the highest fee was always confirmed in the block, while the lower-fee transaction and the legitimate transaction were discarded. The outcome of the attack was solely dependent on the fee rate but not the order of transaction submission.

In the third round, we increased the complexity of the experiment by introducing multiple sniping attempts. In this case, we set up four different attack transactions with varying fee rates (e.g., 9 sats/vB, 20 sats/vB, 35 sats/vB, and 50 sats/vB). We ran the experiment again under these conditions, and as expected, the transaction with the highest fee (50 sats/vB) was always the one included in the block, while the others were removed from the mempool due to conflicts. This third round demonstrated that the snipping attack could consistently be replicated, even when multiple transactions with different fees were used.

%================================================= 
\section{Attacking Adaptability}
\label{sec-applicable}
%================================================= 

We discuss the adaptability of our attack and examine its applicability to various platforms.

\subsection{Platform Requirements}
For a BRC20 snipping attack to be possible, the following platform conditions must be met:

\smallskip
\noindent\textbf{PSBT support.} The platform must utilize or support the creation, signing, and broadcasting of PSBTs. In our attack model, these transactions are broadcasted with incomplete signatures, leaving room for a buyer to sign and complete the transaction.

\smallskip
\noindent\textbf{Unconfirmed transactions in mempool.} The platform must allow transactions to stay in an unconfirmed state long enough for an attacker to access them. This is essential because the snipping attack exploits the fee-based transaction replacement mechanism, where transactions in the mempool can be replaced by those with higher fees before being included in a block. If a platform processes transactions too quickly or confirms them instantly, it would limit the opportunity for an attacker to execute this attack.

\smallskip
\noindent\textbf{Fee transparency.} In many marketplaces, BRC20 tokens are traded with a transparent fee structure, where both the buyer and the seller are aware of the transaction’s proposed fee. If the platform does not provide sufficient visibility into transaction fees or hides this information from participants, it could reduce the effectiveness of the snipping attack, as the attacker would not be able to adjust their fee to outbid the legitimate buyer.

\smallskip
\noindent\textbf{Third-party API support.} Attackers often rely on the ability to query or subscribe to the mempool using third-party APIs or other tools that allow them to observe pending transactions. Platforms that enable API access will make it easier for attackers to spot legitimate BRC20 transactions.

\subsection{Token Requirements}
BRC20 tokens are a specific implementation of token standards built on the Bitcoin, typically through the use of inscriptions that embed JSON-like data directly into satoshis. The success of the snipping attack depends on several features of the BRC20 token standard and how it integrates with Bitcoin’s transaction system. However, the attack’s applicability could vary depending on the particularities of the token standard and its underlying structure.

First, BRC20 tokens, like other Bitcoin-based tokens, rely on Bitcoin’s UTXO model for transaction handling. For the snipping attack to work, the BRC20 token must be able to function within this model, meaning that the tokens must be transferred as part of a valid Bitcoin transaction. This means that token transfers must rely on inputs and outputs in the Bitcoin transaction that can be partially signed, and these signatures must be modifiable by the buyer and the attacker.

Second, a key element in BRC20 token transactions is the embedding of metadata (e.g., token type, amount, operation) in the Bitcoin transaction’s \texttt{witness} or an \texttt{OP\_RETURN} field. This ensures that each token transfer includes a data payload that can be manipulated or read by an attacker looking for transactions with high fees. If a token standard uses different mechanisms for data storage or transaction processing, such as different witness formats or proprietary methods for encoding token data, this could limit the applicability of the attack. Therefore, the attack is well-suited to tokens that are inscribed in the manner of BRC20, where metadata is clearly visible and can be captured in the mempool.

Furthermore, BRC20 tokens, like most Bitcoin transactions, depend on the finality provided by block confirmation. The attack assumes that the target transaction is unconfirmed and still in the mempool, allowing the attacker to insert a higher-fee transaction that replaces the original. This finality characteristic can vary based on how the transactions are managed. For example, if the token platform uses alternative finality mechanisms that do not rely on Bitcoin’s block confirmation times (e.g., through a secondary layer or off-chain settlements), the snipping attack may not be applicable.

\subsection{Applicability to Other Tokens}
Our snipping attack is primarily designed for BRC20 tokens. However, the general principles of the attack could be adapted to other token standards, depending on how they are integrated with Bitcoin’s base layer and whether the platform employs similar transaction mechanisms. We examine the potential for applying the attack to other token systems like Rune in Table~\ref{tab:brc-tokens}.

Rune, another token standard designed for use within the Bitcoin ecosystem, could also be vulnerable to a form of snipping attack. Although Runes also leverage Bitcoin’s core framework of satoshi-level inscription, they adopt a distinct approach compared to BRC20 by splitting the inscription data and functional definitions differently. Runes often separate key state-tracking details from the operational constraints of token transfers, potentially distributing them across multiple inscription contexts. Nevertheless, like any Bitcoin-based inscription method, Runes relies on unspent outputs and transaction finalization through block inclusion. Consequently, if a Rune transaction is broadcast as a partially signed or unconfirmed transaction visible in the mempool, it remains vulnerable to the snipping attack.

\begin{table}[!t]
  \centering
  \caption{Attacking Applicability to Tokens in Various Platforms}  \label{tab:brc-tokens}
      \vspace{-0.2cm}
\renewcommand{\arraystretch}{1} 
  \begin{threeparttable}
 \resizebox{\linewidth}{!}{
  \begin{tabular}{c|c c c | ccccc }
    \toprule
    \multicolumn{1}{c}{\textbf{Tokens}}& 
    \multirow{1}{*}{\textbf{M.Cap}}& 
    \multirow{1}{*}{\textbf{Protocol}} & 
    \multicolumn{1}{c}{\multirow{1}{*}{\textbf{Network}}} & 
    \rotatebox{80}{\textbf{Binance}} & 
    \rotatebox{80}{\textbf{Unisat}} & 
    \rotatebox{80}{\textbf{Magic Eden}} & 
    \rotatebox{80}{\textbf{Gate.io}} & 
    \rotatebox{80}{\textbf{OKX}} \\
    
   \midrule
   
   Ordi & \$711M &  BRC20 & Bitcoin & \cmark & \cmark & \cmark & \cmark & \cmark  \\ 
    
    Sats & \$612M &  BRC20 & Bitcoin & \cmark & \cmark & \xmark & \cmark & \cmark  \\ 
    
    Rats & \$115M &  BRC20 & Bitcoin & \cmark & \cmark & \xmark & \cmark & \cmark  \\ 

    LeverFi & \$74M  &  BRC20 & Bitcoin & \cmark & \cmark & \xmark & \cmark & \cmark  \\ 

    PIZZA & \$61M &  BRC20 & Bitcoin & \cmark & \cmark & \xmark & \cmark & \cmark  \\ 

    WZRD & \$57M &  BRC20 & Bitcoin & \cmark & \cmark & \xmark & \cmark & \cmark  \\ 

    PUPS & \$48M &  BRC20 & Bitcoin & \cmark & \cmark & \xmark & \cmark & \cmark  \\ 

    TEXO & \$46M &  BRC20 & Bitcoin & \cmark & \cmark & \xmark & \cmark & \cmark  \\ 

    Multibit & \$33M &  BRC20 & Bitcoin & \cmark & \cmark & \xmark & \cmark & \cmark  \\ 

    TRAC & \$28M &  BRC20 & Bitcoin & \cmark & \cmark & \xmark & \cmark & \cmark  \\ 

    Atomicals & \$27M &  ARC20 & Bitcoin & \xmark & \xmark & \xmark & \xmark & \xmark  \\ 

    Piin & \$11.27M &  BRC20 & Bitcoin & \cmark & \cmark & \xmark & \cmark & \cmark  \\ 

    Orange & \$8M &  BRC20 & Bitcoin & \cmark & \cmark & \xmark & \cmark & \cmark  \\ 

    BNSx & \$1.22M &  BRC20 & Bitcoin & \cmark & \cmark & \xmark & \cmark & \cmark  \\ 
    
    cats & \$1.19M &  BRC20 & Bitcoin & \cmark & \cmark & \xmark & \cmark & \cmark  \\ 

    Runes & \$541,682 &  Rune & Bitcoin & \cmark & \cmark & \cmark & \cmark & \cmark  \\ 

    \cmidrule{1-6}

    ETHS & \$19M &  ETHS20 & Ethereum & \xmark & \xmark & \xmark & \xmark & \xmark  \\ 

    DOGI & \$17M &  DRC20 & Dogecoin & \xmark & \xmark & \xmark & \xmark & \xmark  \\ 

    Gram & \$9.89M &  TON20 & Arbitrum & \xmark & \xmark & \xmark & \xmark & \xmark  \\ 

    BSCS & \$1.15M &  BSC20 & BNB Chain & \xmark & \xmark & \xmark & \xmark & \xmark  \\ 

    SOLS & \$90,278 &  SPL20 & Solana & \xmark & \xmark & \xmark & \xmark & \xmark  \\ 

    \cmidrule{4-9}
    
    \multicolumn{4}{c}{} & \multicolumn{5}{c}{\textbf{Attacking Applicable?}} \\

     \bottomrule
  \end{tabular}
  }

\begin{tablenotes}[flushleft]
    \footnotesize
    \item \textbf{Binance}: The largest centralized exchange worldwide for trading blockchain tokens. 
    \item \textbf{Unisat}: A user-facing wallet and marketplace specifically tailored for BRC20 tokens.
    \item \textbf{Magic Eden}: A prominent multi-chain NFT and token marketplace. Some tokens are \\\textit{not available for trading} or are not part of the marketplace’s featured offerings.  \\ They only serve as (\textit{NFT collection}). 
    \item \textbf{Gate.io}: A centralized exchange where trading is typically managed internally.
    \item \textbf{OKX}: A centralized exchange offering digital asset trading.
\end{tablenotes}

  \end{threeparttable}
\end{table}

%================================================= 
\section{Mitigation Strategies}
\label{sec-mitigation}
%================================================= 

Mitigating the risk of snipping attacks in a BRC20 marketplace involves measures that either anticipate the possibility of fee-driven replacement or swiftly respond once a transaction is at risk of being overtaken. From our observations, there are two primary strategies that have been implemented in certain platforms (e.g., Magic Eden) to secure token transfers: \emph{Adding Mempool Protection} and \emph{Increasing Fees} after submission. Apart from that, we proposed a more fundamental mitigation that can be pursued at the protocol, termed as \emph{Advanced Fee-Locking Mechanism}. 

\subsection{Adding Mempool Protection}
The first approach centers on a proactive layer of defense termed partial mempool protection. This mechanism operates by automatically broadcasting a series of higher-fee transactions on behalf of the user if the original transaction becomes vulnerable to being replaced or sniped. Once the user opts in, they can specify a maximum fee multiplier relative to the prevailing network fee rate. When proceeding to make a purchase, three transaction signatures are typically required. The first transaction uses the base network fee (for instance, 9 sats/vB), while the second and third pre-authorize higher fee rates that can be broadcast if the initial attempt is supplanted. Concretely, as shown in Fig.\ref{fig:add-mempool}, a user might sign an initial purchase at 9 sats/vB, then authorize the platform to escalate the fee to 95 sats/vB if the first transaction is sniped, and finally approve a peak fee of 180 sats/vB should the second transaction also fail. This tiered arrangement enables the user to maintain control of the maximum fee while still protecting against multiple rounds of replacement by an adversary.

\subsection{Increasing Fees after Submission}
A second mitigation strategy allows users to manually heighten their transaction fees once the purchase is pending in the mempool (shown in Fig.\ref{fig:increase-fee}). After submitting an order, the user can track its status on a dedicated “Pending Orders” page. If the transaction remains valid and unconfirmed, a “Top Fee” status indicates that it has not been overridden; however, if the status shows “Getting Replaced,” this signals that a snipping attempt may be underway, prompting the user to confirm a higher fee rate. By selecting the “Increase Fee” option, users can override their original fee with a new, higher rate exceeding any recommended minimum. Should the mempool or the platform detect that the user’s transaction is still active, this re-issuance with a higher fee can successfully restore priority and secure the purchase. In effect, this second approach grants users direct control over re-broadcasting their transaction at an elevated fee without relying on a prearranged tiered schedule.

\subsection{Advanced Fee-Locking Mechanism}
While adding mempool protection and increasing fees after submission offer user-level responses to snipping threats, we propose \emph{advanced fee-locking mechanism} at the protocol to prevent external parties from arbitrarily escalating fees on PSBTs. At its core, our approach involves cryptographic commitments that embed a maximum allowable fee within the partially signed data, ensuring that no external actor can alter that fee.

\smallskip
\noindent\textbf{Commitment scheme.} 
A commitment-based design requires users to commit to a maximum fee (e.g., \(f_{\mathrm{max}}\)) when they first assemble the PSBT. This commitment might look like:
\[
\mathrm{commitFee}(f_{\mathrm{max}}) = \mathcal{H}\bigl(\text{encode}(f_{\mathrm{max}}) \Vert \text{nonce}\bigr),
\]
where \(\mathcal{H}\) is a secure hash function, and the random nonce conceals the exact value of the user’s maximum fee until finalization. Once the user has committed to this fee lock, the transaction’s subsequent steps must not exceed \(f_{\mathrm{max}}\), or else the commitment becomes invalid. When the transaction is fully signed and ready for broadcast, the user reveals the nonce and the specific fee level, ensuring all nodes can verify that the fee remains within the committed bound.

\smallskip
\noindent\textbf{Mechanics of finalization and broadcast.} 
In practice, the scheme would proceed as follows: 
(i) The user computes a commitment to \(f_{\mathrm{max}}\) and embeds it into an extra output script or the PSBT witness field;
(ii) The user partially signs the transaction, ensuring that the script logic rejects any attempt to set fees above \(f_{\mathrm{max}}\);
(iii) The final signature phase reveals the pre-image of \(\mathrm{commitFee}(f_{\mathrm{max}})\), thereby binding the on-chain transaction to a specific actual fee; 
(iv) The broadcast transaction includes this revealed fee, allowing miners to validate that the transaction satisfies its original commitment.

%================================================= 
\section{Related Work}
%================================================= 

\noindent\textbf{Transaction replacement.}
Transaction replacement allows for substituting a previously sent but unconfirmed transaction with a new one. In Bitcoin, this is commonly done using replace-by-fee (RBF, outlined in BIP-125~\cite{bip125}. RBF adds a layer of strategy to adjusting transaction orders~\cite{li2023transaction}, yet it necessitates cautious handling to sidestep security risks such as pinning attacks~\cite{qi2024brc20,txpinHTLC}. Importantly, RBF is initiated by the asset owners themselves.

Transaction replacement can also be initiated by miners, a practice known as miner-extracted value (MEV) \cite{yang2024sok,qin2022quantifying,daian2020flash}. MEV is prevalent in the Ethereum ecosystem and has spread to other blockchain platforms. This practice can disadvantage users (e.g., reorg attacks) \cite{daian2020flash,heimbach2022sok,eskandari2020sok}), leading to potential losses and compromising the fairness of transactions \cite{li2023transaction,kelkar2020order}.

\smallskip
\noindent\textbf{Blockchain sniper.} A sniper bot is an automated tool designed to outpace all participants in purchasing tokens, leveraging blockchain mempool activity for malicious purposes. These activities include targeting short-lived tokens (e.g., 1-day rug pulls) on EVM token-issuing platforms~\cite{cernera2023token,cernera2023ready,cernera2024warfare} and exploiting MEV arbitrage opportunities in DeFi protocols~\cite{eskandari2020sok,heimbach2022sok,yang2024sok}. Instead of directly deploying such bots, we adopt a similar strategy by monitoring the mempool to intercept the initial transaction containing a buy request, replacing it with a falsified one to disrupt its execution. Our attack contrasts with mainstream frontrunning attacks~\cite{zhou2021high,daian2020flash,wang2022exploring}, which can either manipulate transactions, miners’ blocks, or miner rotations.

\smallskip
\noindent\textbf{BRC20/inscription.}  The concepts of BRC20 and inscriptions were widely introduced~\cite{binance1,binance2,binance3,li2024bitcoin} and explored across various dimensions, including social sentiment~\cite{wang2023understanding}, on-chain fee fluctuations~\cite{bertucci2024bitcoin}, off-chain indexers~\cite{wen2024modular}, component enhancements~\cite{wang2024bridging}, and security analyses~\cite{qi2024brc20,liu2024push}.
Inscirption-based techniques can empower Bitcoin with more functionalities to further establish layer-two solutions~\cite{qi2024sok,bitvm,linus2024bitvm2}. At the moment, they remain in infancy.

\smallskip
\noindent\textbf{Mempool-related attacks} (Table~\ref{tab:attack_comparison}). Blockchain mempool attacks include manipulating or exploiting the unconfirmed transactions within the mempool to achieve benefits. Mempool attacks include transaction malleability attack~\cite{decker2014bitcoin,andrychowicz2015malleability} (altering the digital signature of a transaction to change its ID before confirmation), frontrunning or sandwitching attack~\cite{daian2020flash,zhou2021high,zhang2022frontrunning,wang2022exploring} (placing a transaction with a higher/lower fee to precede/follow another transaction for profit), double spend attack~\cite{negy2020selfish,feng2019selfish,eyal2018majority} (sending two conflicting transactions or braches to favor one over the other), timejacking attack~\cite{zhang2023time} (manipulating node timestamps to affect unconfirmed transaction timing), imitation attack~\cite{qin2023blockchain} (replicating a victim's transaction or contract logic for financial gain) and pinning attacks~\cite{qi2024brc20,txpinHTLC}, DDos attacks~\cite{wu2020survive,saad2019shocking,saad2019mempool,vasek2014empirical} (overloading the mempool to disrupt transaction processing and mempool functionality).

%=================================================
\section{Conclusion}
\label{sec-conclusion}
%================================================

%In this paper, we investigated the emerging threat of mempool sniping in BRC20 trading, uncovering how the partial reliance on PSBTs and fee-based prioritization can enable malicious actors to front-run legitimate transactions. Our controlled environment experiment substantiated how trivial it can be for an attacker with minimal resources to invalidate a buyer’s transaction simply by paying a higher fee. We believe our examination of the BRC20 Snipping Attack will serve as a foundation for future protocol-level improvements and marketplace designs, ultimately contributing to a safer environment for trading on Bitcoin.

In this paper, we present the BRC20 sniping attack, a growing threat to the open market for BRC20 tokens. BRC20 open marketplace relies on partially signed transactions (PSBT), which we exploit by manipulating transaction priorities. By allocating unfair fees, malicious actors can front-run legitimate transactions. Our attack is applicable to most mainstream BRC20 trading platforms.

We implemented and evaluated the attack. Experimental results show that even attackers with minimal resources can invalidate a buyer's transaction by simply paying a higher fee. We responsibly report our attack to related platforms. We hope this security analysis of BRC20 will lay the groundwork for protocol-level improvements and better marketplace designs in the Bitcoin ecosystem.

\smallskip
\noindent\textbf{Acknowledgement.}
We would like to thank Yingjie Zhao (Minzu University of China) for his valuable contributions to the experimental portion of this research. His assistance was instrumental in refining the methodology presented in this paper.

\begin{table*}[!h]
\renewcommand{\arraystretch}{1.1}
\centering
\caption{Abbreviations (i.e., \textcolor{teal}{type.id}) in our attack}
\resizebox{\textwidth}{!}{
\begin{tabular}{cc|cc|c}
\toprule
\multicolumn{2}{c}{\textbf{Type}}  & \multicolumn{1}{c}{\textbf{ID}}  & \multicolumn{1}{c}{\textbf{Full ID}} & \textbf{Notes} \\
\midrule
\multirow{5}{*}{\rotatebox{90}{Address}}   & \textcolor{teal}{adr.} & \textcolor{teal}{0dd8} & bcrt1qkckuparv0d8nxadvzxgl0m0fmsn0ym6xfq0dd8 & Seller's wallet address for selling BRC20 tokens \\
 & \textcolor{teal}{adr.} & \textcolor{teal}{qa8r} & bcrt1qackm4gsk5szqmylr0k5fq2n6v9jugz4473qa8r & Buyer's wallet address for receiving BRC20 tokens\\ 
 & \textcolor{teal}{adr.} & \textcolor{teal}{v6r9} & bcrt1qtpq478leuks3wcg9zgr7rqhz6enszpsmt7v6r9 & Attacker's wallet address for snipping BRC20 tokens with \textit{high} fee \\
 & \textcolor{teal}{adr.} & \textcolor{teal}{xffg} & bcrt1quee4tw5xwxw236yuw8zyuw83r36dtnztsuxffg & Attacker's wallet address for snipping BRC20 tokens with \textit{low} fee\\
 & \textcolor{teal}{adr.} & \textcolor{teal}{tdks} & bcrt1qt30mvqww2cq9nvq4zlefj0l330y4k2ulqtdyks & Miner's wallet address for mining block and receive mining reward\\ 
 
 \cmidrule{1-3} 
 
\multirow{5}{*}{\rotatebox{90}{Transaction}}  & \textcolor{teal}{txid} & \textcolor{teal}{240c} & 1d153468fbb27178ed84709fcdb22d6fef4140d67fa887166451b63a24a9240c & Transaction of the \textit{seller} linked to an UTXO for \textit{deploy}\\

  & \textcolor{teal}{txid} & \textcolor{teal}{641d} & 3aad7f5f7d3e39bf7a98741550de767ff191d1ebfe9326177825890ec92d641d & Transaction of the \textit{seller} linked to an UTXO for \textit{createpsbt}\\

 &  \textcolor{teal}{txid} & \textcolor{teal}{b2ea} & 6ec491e4928d9b6af81b0ebd5040e1a44834f2576e1085ecb79d5fddc4fbb2ea & Transaction of the \textit{buyer} linked to an UTXO for \textit{createpsbt} \\
 
 &  \textcolor{teal}{txid} & \textcolor{teal}{c936} & 1d269e83d7a415197038372196ad0c19fc380861b51e788cd3ec0fced777c936 & Transaction from the buyer broadcasts the completed PSBT \\

&  \textcolor{teal}{txid} & \textcolor{teal}{f0bb} & e5be4872b8dd95369df0d15846e5099b5f20662e9809e65348151a9a257af0bb & Transaction from the attacker broadcasts the completed PSBT (\textit{high fee}) \\

&  \textcolor{teal}{txid} & \textcolor{teal}{2323} & 3abe4872b8dd95369df0d15846e5099b5f20662e9809e65342323aodsa2323 & Transaction from the attacker broadcasts the completed PSBT (\textit{low fee}) \\

 \cmidrule{1-3} 
 
 \multirow{2}{*}{\rotatebox{90}{Block}} & \textcolor{teal}{block} & \textcolor{teal}{a2e2} & 6ffa7cb6d94727f6a619d4a33c60f0f65b8b7f72dde3e5fd3fd9e94753c6a2e2 & Block containing a BRC20 deploy transaction \\
 
 & \textcolor{teal}{block} & \textcolor{teal}{d848} & 18be98d75b24c4f03e58f17bb9adf098b96c02d39456b74e9217ff4bcbd8d848 & Block containing brc20 transfer pbst transaction \\

 \cmidrule{1-3} 
 \multirow{5}{*}{\rotatebox{90}{Data Hex}} & \textcolor{teal}{hex.} & \textcolor{teal}{307d} & 7b22703a20226272632d3230222c20226f70223a20226465706c6...13030307d & Hexadecimal-encoded string of BRC20 token deploy data \\

 & \textcolor{teal}{hex.} & \textcolor{teal}{227d} & 7b22703a226272632d3230222c226f70223a227472616e7366657...03030227d & Hexadecimal-encoded string of BRC20 token transfer data\\

 & \textcolor{teal}{hex.} & \textcolor{teal}{e140} & 02000000000101eab2fbc4dd5f9db7ec85106e57f23448a4e140...8200000000 & Buyer created completed psbt hex-encoded string \\

  & \textcolor{teal}{hex.} & \textcolor{teal}{ec85} & 02000000000101eab2fbc4dd5f9db7ec85...15b96648c1ea8ba588200000000 & Attacker created completed psbt hex-encoded string (with \textit{high} fee) \\

  & \textcolor{teal}{hex.} & \textcolor{teal}{4050} & 02000000000101eab2fbc4dd5f9db7ec85106e57f23448a4e14050...0000000 & Attacker created completed psbt hex-encoded string (with \textit{low} fee) \\

 \cmidrule{1-3} 
 
 \multirow{4}{*}{\rotatebox{90}{PSBT}} & \textcolor{teal}{psbt} & \textcolor{teal}{r85P} & cHNidP8BAHYCAAAAAR1kLckOiSV4FyaT/uvRkfF/dt5QFXSYer85P...0AAAAAAAAA & Seller created psbt base64-encoded string \\
 
& \textcolor{teal}{psbt} & \textcolor{teal}{+Gqb} & cHNidP8BALQCAAAAAeqy+8TdX5237IUQblfyNEik4UBQvQ4b+Gqb...AAAgAAAAAA & Buyer created psbt base64-encoded string \\

& \textcolor{teal}{psbt} & \textcolor{teal}{UBQv} & cHNidP8BAJUCAAAAAeqy+8TdX5237IUQblfyNEik4UBQv...AAIAAAAAAAgAAAAAA & Attacker created psbt base64-encoded string (with \textit{high} fee) \\

& \textcolor{teal}{psbt} & \textcolor{teal}{b+Gq} & cHNidP8BALQCAAAAAeqy+8TdX5237IUQblfyNEik4UBQvQ4b+Gq...AAABAAAAAAA= & Attacker created psbt base64-encoded string (with \textit{low} fee) \\
 
\bottomrule
\end{tabular}
}
\end{table*}

%================================================
\bibliographystyle{unsrt}
\bibliography{bib.bib}
%================================================

\appendix

\begin{figure}[t]
    \centering
    % First subfigure
    \subfigure[Adding Mempool Protection Mechanism]{\label{fig:add-mempool}
        \includegraphics[width=\linewidth]{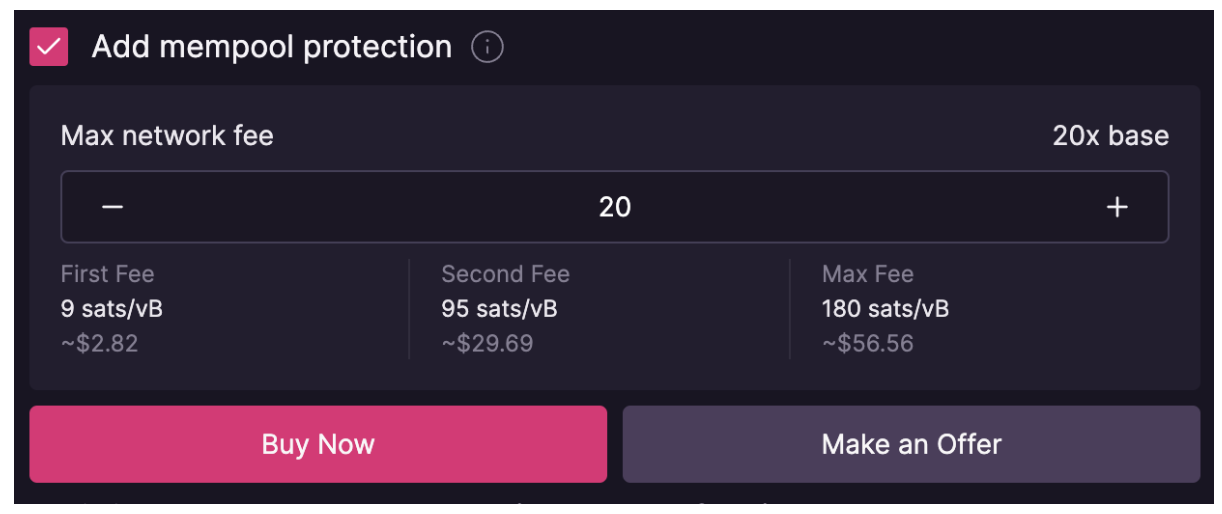}
    }
 %   \vspace{-0.7in}
    \subfigure[Increasing Fees After Submission]{\label{fig:increase-fee}
        \includegraphics[width=\linewidth]{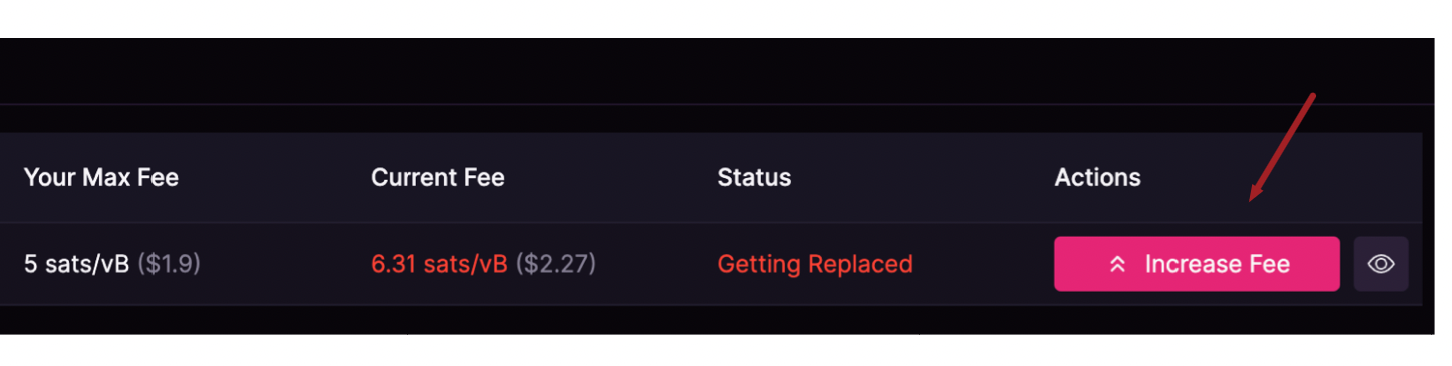}
    }
    \vspace{-0.15in}
    \caption{Screenshots for Mitigation Strategies}
   % \vspace{-0.15in}
\end{figure}

\section{Bitcoin Historical Prices}

\begin{figure}[!]
    \centering
    \includegraphics[width=\linewidth]{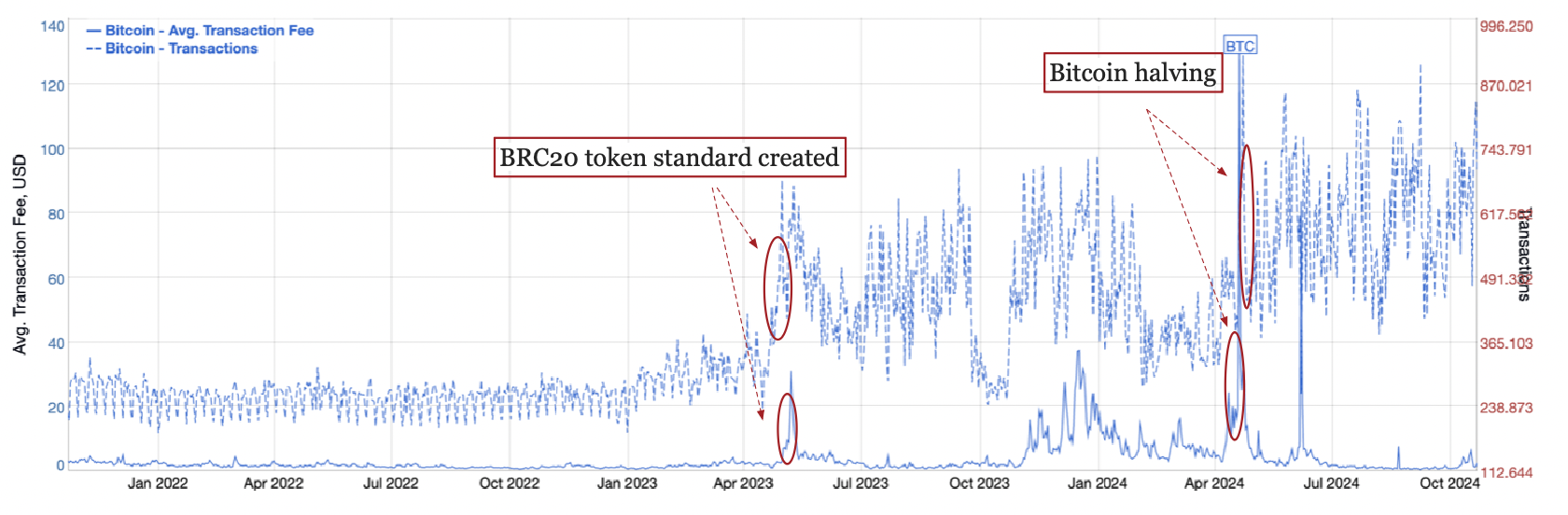}
    \caption{Bitcoin Avg. Transaction Fee vs. Transactions Per Day historical chart (data source: \cite{bitinfocharts24})}
    \label{fig:tx-fee}
    \vspace{-0.25in}
\end{figure}

Fig.\ref{fig:tx-fee} illustrates the correlation between the average Bitcoin transaction fee (in USD) and the number of transactions per day from 2022 to 2024. We observe that the average transaction fee exhibits fluctuations, reflecting changes in network demand and activity. The number of daily transactions shows periodic spikes, often coinciding with specific events or external factors influencing the Bitcoin network.

In early 2023, the creation of the BRC20 token standard marks a significant event, as highlighted in the first red circle. This innovation introduced a new layer of activity to the Bitcoin blockchain by enabling token issuance directly on-chain. The subsequent increase in activity led to network congestion, as seen in the spike in transaction fees and daily transactions during this period. Higher fees result from users bidding to have their transactions prioritized in the increasingly crowded mempool.

Later in the chart, the second red circle aligns with Bitcoin’s halving event. The halving reduced miner rewards, directly impacting miner incentives. This likely increased transaction fees temporarily, as miners prioritized high-fee transactions to compensate for reduced block subsidies. This event also correlated with a spike in daily transactions, possibly reflecting increased trading activity and network adjustments around the halving.

\begin{comment}

\section{Additional Screenshots}

\qw{remove, but update context in corresponding sections}

\begin{figure}[h]
    \centering
    \includegraphics[width=\linewidth]{img/mempool.png}
    \caption{Mempool Observation After the Snipping Attempts}
    \label{fig:mempool_three_tx}
    \vspace{-0.15in}
\end{figure}

\begin{figure}[h]
    \centering
    \includegraphics[width=\linewidth]{img/token_amount.png}
    \caption{Token Balances Tracking}
    \label{fig:block}
    \vspace{-0.15in}
\end{figure}

\end{comment}

\end{document}